\documentclass{article}
\usepackage{times}
\usepackage{graphicx}

\usepackage{makecell}

\usepackage{adjustbox}

\usepackage{amsthm}
\usepackage{amsmath}
\usepackage{amsfonts}
\usepackage{amssymb}
\usepackage{units}
\usepackage{bbm}

\usepackage{natbib}

\usepackage{algorithm}
\usepackage{algorithmic}

\usepackage{hyperref}

\theoremstyle{remark}

\usepackage{wrapfig}
\usepackage{subfigure} 
\usepackage{float}

\usepackage[margin=1.3in]{geometry}

\usepackage{mathtools}
\usepackage{thmtools, thm-restate} 

\usepackage{pdfpages}

\title{Suboptimal and trait-like reinforcement learning strategies correlate with midbrain encoding of prediction errors}

\author{Liran Szlak, Kristoffer Aberg, Rony Paz}

\begin{document}

\maketitle

\begin{abstract}
During probabilistic learning organisms often apply a sub-optimal ‘probability-matching’ strategy, where selection rates match reward probabilities, rather than engaging in the optimal ‘maximization’ strategy, where the option with the highest reward probability is always selected. Despite decades of research, the mechanisms contributing to probability-matching are still under debate, and particularly noteworthy is that no differences between probability-matching and maximization strategies have been reported at the level of the brain. Here, we provide theoretical proof for a computational model that explains the complete range of behaviors between pure maximization and pure probability-matching. Fitting this model to behavior of 60 participants performing a probabilistic reinforcement learning task during fMRI scanning confirmed the model-derived prediction that probability-matching relates to an increased integration of negative outcomes during learning, as indicated by a stronger coupling between midbrain BOLD signal and negative prediction errors. Because the degree of probability-matching was consistent within an individual across nine different conditions, our results further suggest that the tendency to express a particular learning strategy is a trait-like feature of an individual.
\end{abstract}

\section{Introduction}
How we perceive the world depends on the information we are exposed to and what we learn from it, both which are determined by the learning strategy we apply. Experimentally, an individuals' learning strategy can be assessed via the multi-armed bandit task, in which engaging with different arms (i.e. the available options in the world) provides positive and negative feedback with different probabilities. When reward contingencies are stationary (e.g. one option always provides positive feedback with a higher probability than the other options), the optimal strategy is pure ‘maximization’, in which an individual always exploit the option with the highest probability of yielding positive feedback. However, people often apply a sub-optimal ‘probability-matching’ strategy, in which the frequency of selecting an option is proportional to its probability of receiving positive feedback \citep{gaissmaier2008smart, morse1960probability, norman1966probability, rivas2013probability}. \\
	
Different higher order cognitive processes have been proposed to account for human probability-matching (for a review, see~\cite{vulkan2000economist}). Some of these include being unaware of different strategies \citep{newell2013probability}, assigning a higher utility for predicting infrequent outcomes \citep{brackbill1962supplementary}, trying to find sequential patterns \citep{gaissmaier2008smart}, or boredom \citep{shanks2002re}. However, probability-matching has also been observed in non-human animals \citep{gallistel2007matching, herrnstein1961relative, herrnstein1990rational, sugrue2004matching}, which therefore questions to what extent higher order cognitive processes are necessary to express probability-matching. Indeed, while some machine learning models have suggested that learning strategies relate to decision making biases \citep{bubeck2012regret, norman1966probability,rivas2013probability, sakai2008does}, pure probability-matching can also be explained by the simple learning rule in which only the most recent outcome is used to guide the next decision, i.e. the win-stay/lose-shift strategy \citep{gaissmaier2008smart}. To clarify, an option whose selection results in a positive outcome is repeated (win-stay strategy) while options providing negative outcomes are abandoned (lose-shift strategy). Notably, win-stay/lose-shift strategies can be described via standard reinforcement learning models which posit that learning occurs when there is a mismatch between a predicted and an actual outcome, the so called prediction error (PE), scaled by a learning rate \citep{sutton2018reinforcement}. The learning rate determines how quickly behavior changes after a PE has been elicited, and a win-stay/lose-shift strategy corresponds to setting the learning rate to its maximum value (the learning rate is usually bounded between 0 and 1). However, learning rates fitted to human behavior in probabilistic reinforcement learning tasks are often much smaller (e.g. closer to 0 than to 1), and may also differ between learning rates that scale positive and negative PEs (i.e. outcomes that are respectively more positive and negative than predicted; \citep{carl2016linking, dorfman2019causal, gershman2015learning, lefebvre2017behavioural}. Therefore, while a win-stay/lose-shift strategy is capable of explaining probability-matching in theory \citep{gaissmaier2008smart}, it is unclear to what extent humans actually apply such a strategy, and in particular how maximizing behaviors relate to deviations from this strategy (but see \citep{da2017exploration}). Furthermore, few studies have reported differences in the neuronal correlates of maximization versus probability-matching beyond the level of a synapse \citep{loewenstein2006operant}. However, correlates of PEs have been consistently reported in midbrain dopamine neurons \citep{schultz2016dopamine}, with related neuronal correlates observed using fMRI \citep{aberg2015hemispheric, d2008bold}. Additionally, individuals that were able to learn a difficult reinforcement learning task showed stronger correlations between model-derived PEs and BOLD signal in the striatum, as compared to individuals that failed to learn \citep{schonberg2007reinforcement}, and biases in learning from different types of feedback were associated with stronger encoding of their respective PEs in striatal and midbrain regions \citep{aberg2015hemispheric, aberg2016left}. Therefore, if the difference between probability-matching and maximization strategies relates to learning biases, a straightforward prediction is that those differences should be evident also in the neuronal encoding of PEs. \\
	
To test these hypotheses, we first provide theoretical proof for a linear reinforcement learning model that gives rise to the full range of strategies from full maximization to full probability-matching. In brief, this model suggests that an increased expression of a probability-matching strategy relates to an increased integration of negative outcomes during learning. The model was then fitted to behavioral data in 77 participants performing a reinforcement learning task (60 of those performed the task during fMRI scanning). Participants expressed a range of learning strategies between full maximization and full probability-matching, and these strategies were found to be consistent within an individual across nine different task conditions. Finally, confirming the critical model-prediction, fMRI data showed that individuals expressing more probability-matching displayed stronger correlations between midbrain BOLD signals and negative PEs (i.e. PEs elicited by negative outcomes). \\ 
	
In summary, our results indicate that reinforcement learning biases contribute to the trait-like expression of a particular learning strategy. Our results also provide a clear description of the mechanisms contribution to probability-matching, which is well-anchored in reinforcement learning theory, and which does not require the involvement of higher order cognitive processes (but which may interact with them). Beyond learning theory, our results may inform psychopathology associated with both abnormal learning- and decision-making behaviors, and it is tempting to suggest that dysfunctional and destructive learning strategies could be targeted as part of preventive and therapeutic approaches to combat mental illness.
\section{Methods}
\subsection{Participants}
Seventy-seven participants took part in the experiment. Sixty of the participants (Mean age $\pm$ SEM: 26.05 $\pm$ 0.44; 35 females) performed the task inside the MRI scanner, and are included both in the behavioral and fMRI results. Seventeen different participants performed the task outside the MRI scanner and are included in the behavioral results. Unfortunately, the demographic data for these participants were lost, but they were recruited from the same pool of participants as those of the fMRI study. All participants signed informed consent and the study was conducted in accordance with the ethical review board of the Weizmann Institute of Science and the declaration of Helsinki. Six participants were excluded from all analyses due to having low model-fitted learning rates ($<0.05$) in the majority of conditions. We note that including these six participants does not change the analysis results or the conclusions. 
\subsection{Probabilistic Reinforcement Learning Task}
\begin{figure*}
\includegraphics[scale=0.8]{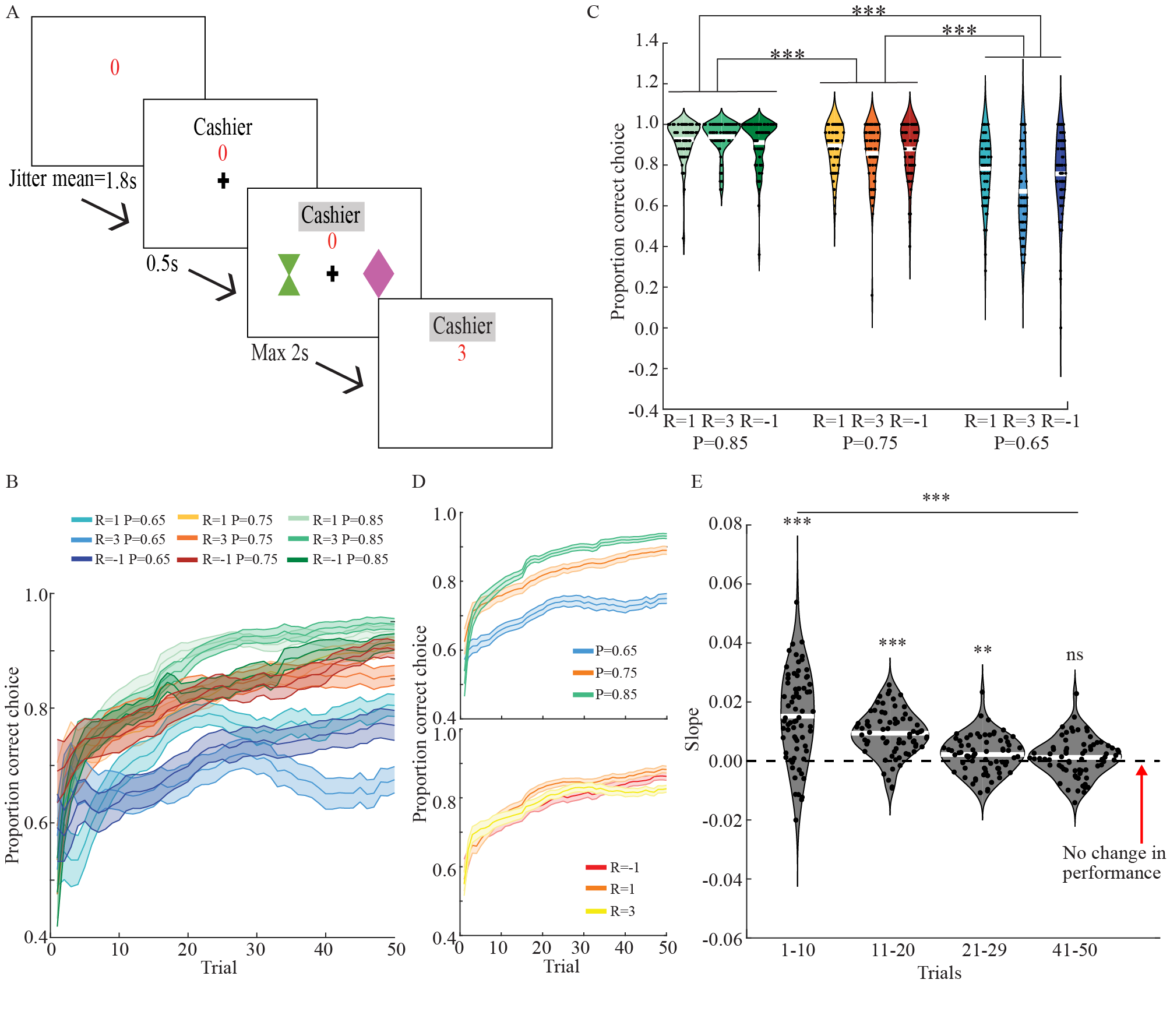}
\caption{Learning task and performance. A. Graphic description of a trial in the task. In this example, the choice resulted in a gain of 3 points as the Cashier increased from 0 to 3. B. The mean proportion of correct choice in each condition as a function of trial. The proportion of correct choice in each condition was calculated over a sliding window of 15 trials. The shaded areas represent SEM. R=outcome Type, P=outcome Probability.  C. The proportion of correct choice in the last 25 trials in each condition. The mean across participants is shown as a white horizontal line inside each violin plot, while each dot represents a participant. D. The average proportion of correct choice as a function of trial collapsed across outcome Probability (top) and outcome Type (bottom). R=outcome Type, P=outcome Probability. The shaded areas represent SEM. The proportion of correct choice in each condition was calculated over a sliding window of 15 trials. E. Distribution of learning slopes calculated during different parts of the learning. The slopes were calculated using linear regression over the proportion of correct choice. Learning occurs during the first ten trials, as indicated by mean slope values significantly different from 0, while no learning occurs during the last ten trials, as indicated by mean slope values centered around 0. White horizontal lines indicate the group mean while the dots indicate an individuals' mean slope value. ***$p<0.001$, **$p<0.01$, ns not significant ($p>0.05$). }
\label{fig:probMatchFig1}
\end{figure*}

In each trial, participants chose one of two shapes (Figure~\ref{fig:probMatchFig1}A), and the outcome depended on the outcome Probability assigned to the selected shape and the condition in which it was presented (Table \ref{table:t1}). Each pair of shapes was presented for 50 continuous trials in one of nine different conditions. Accordingly, participants could learn the ‘best’ shape in each pair of shapes by sampling their respectively outcome probabilities, i.e. one shape in each pair had a higher probability of receiving the positive outcome. The outcome Probability assigned to each shape was static throughout the experiment and the outcome probabilities of the two shapes summed to 1(i.e. if one shape had an outcome Probability of P=0.85, the other shape in the pair had an outcome Probability of 1-P=0.15). The positive outcome in Conditions 7-9 was the value 0 (i.e. no loss occurred) while the value 0 was the negative outcome in Conditions 1-6 (i.e. no gain occurred). The goal of participants was to maximize the number of points accumulated during the game, where every 10 points added 1 NIS to a monetary bonus provided at the end of the experiment. The outcome Probabilities (P) and the outcome Types (R) in each of the nine conditions are shown in Table \ref{table:t1}.

Each trial started with the accumulated number of points displayed for a jittered amount of time (Figure~\ref{fig:probMatchFig1}A; drawn from a truncated negative exponential distribution with a mean of 1.8s). A fixation cross was then presented for 0.5s, directly followed by the presentation of two shapes. Participants had two seconds to choose one of the two shapes, and if no shape was chosen within this time a message appeared on the screen and the trial was lost. Outcome feedback consisted of changing the number of accumulated points displayed on the screen, along with a specific sound for reward/loss. A failure to respond resulted in a negative outcome (i.e. 0 or -1, depending on the condition; Table \ref{table:t1}). 
\begin{table}[h]
\begin{center}
 \begin{tabular}{||c c c||} 
 \hline
 Condition & outcome Probability (P) & outcome Type (R) \\ [0.5ex] 
 \hline\hline
 1 & 0.85 & 1 \\ 
 \hline
 2 & 0.75 & 1 \\
 \hline
 3 & 0.65 & 1 \\
 \hline
 4 & 0.85 & 3 \\
 \hline
 5 & 0.75 & 3 \\ [1ex] 
 \hline
  6 & 0.65 & 3 \\ [1ex] 
 \hline
  7 & 0.85 & -1 \\ [1ex] 
 \hline
  8 & 0.75 & -1 \\ [1ex] 
 \hline
  9 & 0.65 & -1 \\ [1ex] 
 \hline
\end{tabular}
  \caption{The different outcome Probabilities (P) and outcome Types (R) in each condition. Observe that the ‘other’ outcome in each condition was 0. In other words, a 0 outcome was the positive outcome in Conditions 7-9, while it was the negative outcome in Conditions 1-6.}
\label{table:t1}
\end{center}
\end{table}

\subsection{Computational Model}
The K-armed bandit problem is formulated by a sequence of random variables, $X_{i,n}$ where $1 \leq  i \leq K$ indicates the shape number and $n\geq 1$ indicates the trial number. In each trial, the learner selects a shape and receives an outcome Type determined by the outcome Probability assigned to the selected shape. We note here that the outcomes are assumed to be binary (i.e. you either receive an outcome or you receive a 0; see Table!\ref{table:t1}). Next, we defined three different learning models. For a pure ‘Maximization’ model and a pure ‘Probability-matching’ model we prove theoretically that behavior converges to one of two extreme strategies – a maximization strategy, where the learner converges to always (i.e. with a probability of $1.0$) choose the ‘best’ shape, and a probability matching strategy, in which the learner converges to choosing each shape with a probability that is proportional to the outcome probability of that shape. A third ‘Combined’ model was then defined in which behavior may converge to any range of these strategies (i.e. from full maximization to full probability matching) via linear interpolation between the learning strategies of the Maximization and Probability-matching models. Proofs of convergence for all three models can be found in Appendix~\ref{sec:appendixA}. \\

We now formally define the models, starting with notations. We define the random variables: \\
$A_{i,n} = \begin{cases} 1 & \mbox{ learner chose shape i at trial n } \\ 
0 & \mbox{ otherwise} 
\end{cases}$. \\
$X_{i,n} = \begin{cases} 1 & \mbox{ outcome recieved for shape i at trial n } \\ 
0 & \mbox{ otherwise} 
\end{cases}$. \\
We will also use the notation: $A_{j(n),n}$ and $X_{j(n),n}$ where $j(n)$ indicates the shape chosen at trial n.\\
The expected reward of shape $i$ is:
$ \mu_{i} = \mathbb{E}[\mbox{reward in shape i}] $ and: $ \mu^\ast = \underset{1 \leq i \leq k}{max}\{\mu_{i}\} $ \\
We are considering at the setting in which $\forall i\in[1,K]: \mu_i \geq 0$ and that $\underset{t=0}{\overset{n}{\sum}} p_i = 1$ where $p_i$ is the outcome probability for shape $i$. \\
A policy, or choice strategy, A, is an algorithm that chooses the next machine to play based on past trials, that consist of past choices and outcomes. \\
The "maximization" strategy, which we denote $ S_{max} $, will always choose the shape with the highest winning probability and thus have an expected profit of:\\
 \begin{align*}
 \mathbb{E}[profit(S_{max})] = \underset{t=0}{\overset{n}{\sum}} \mathbb{E}[X_{max,t}]=\mu^{\ast}\cdot n 
\end{align*}
A "probability matching" strategy, which we denote $ S_{prob} $, will choose each shape proportional to its underlying distribution, thus yielding an expected profit of:
 \begin{align*}
 \mathbb{E}[profit(S_{prob})] = \underset{t=0}{\overset{n}{\sum}} \mathbb{E}[A_{j(t),t}\cdot X_{j(t),t}]\underset{\mbox{per trial, choice and reward are independent of each other}}{\underbrace{=}}
 \end{align*}
 \begin{align*}
\underset{t=0}{\overset{n}{\sum}} \mathbb{E}[A_{j(t),t}]\cdot \mathbb{E}[X_{j(t),t}]= \underset{t=0}{\overset{n}{\sum}} \underset{i=1}{\overset{k}{\sum}} \mathbb{E}[A_{i,t}] \cdot \mathbb{E}[X_{i,t}] = \underset{t=0}{\overset{n}{\sum}} \underset{i=1}{\overset{k}{\sum}} \mu_i \cdot \mu_i = n \cdot \underset{i=1}{\overset{k}{\sum}} \mu_i^2
 \end{align*}
We compare the expected profit of the two strategies and see that the maximization strategy yields higher expected profit than the probability matching strategy:
\begin{align*}
\mathbb{E}[profit(S_{prob})] = n \cdot \underset{i=1}{\overset{k}{\sum}} \mu_i^2 < n \cdot \underset{i=1}{\overset{k}{\sum}} \mu_i \cdot \mu^{\ast} = n \cdot \mu^{\ast} \cdot \underset{i=1}{\overset{k}{\sum}} \mu_i = n \cdot \mu^{\ast} = \mathbb{E}[profit(S_{max})]
\end{align*}
Accordingly, the probability-matching strategy is sub-optimal in a static environment. \\

In all models, we distinguish between the full information scenario and the partial information scenario. In the full information scenario, at trial n the learner chooses a shape $j(n)$. Following the choice, the learner receives an outcome Type based on the outcome Probability of the chosen shape, i.e. $X_{j(n),n}$, as well as the outcomes sampled from all other shapes. Thus, this scenario is referred to as "full information". In the partial information scenario, however, the learner receives only the outcome of the chosen shape. We will refer here to both settings, and discuss the differences and similarities between the two settings.
We are now ready to define the three models.
\paragraph{Modeling a probability-matching strategy}
This is a simple linear reinforcement model. We show in Appendix~\ref{sec:appendixA} that this model converges to a probability-matching strategy.
Let $ p_{i,n+1} $ be the probability of the learner to choose shape $i$ in trial $n+1$, i.e. $ p_{i,n+1}=pr(A_{i,n+1}) $.  
In the full information scenario, the choice probabilities are updated from trial $n$ to trial $n+1$ by the following rule:\\
for $ i=1,2...,K $ and $ 0 \leq \alpha \leq 1:$ \\
$ p_{i,n+1}=\begin{cases}
p_{i,n}+\alpha \cdot(1-p_{i,n}) & if\,\, X_{i,n}=1\\
p_{i,n}-\alpha \cdot p_{i,n} & otherwise
\end{cases}
$ \\
$\alpha$ is a learning rate that determines the rate with which the choice probabilities are updated. We note that in all our models the sum over probabilities to choose each shape sum up to 1 at every iteration. To show this, imagine that for iteration n we have that $\sum_{i=1}^K p_{i,n} = 1$, then, at iteration $n+1$, assuming, without loss of generality, that shape 1 is the shape that won, we have that:
$p_{1,n} + \alpha\cdot(1-p_{1,n}) + \underset{i=2}{\overset{k}{\sum}} p_{i,n} - \alpha \cdot p_{i,n} = \underset{i=1}{\overset{k}{\sum}} p_{i,n} + \alpha - \alpha \cdot \underset{i=1}{\overset{k}{\sum}} p_{i,n} = 1 + \alpha - \alpha = 1 $

This means that at every iteration, we have that $p_{i,n}$ represent a probability distribution.
Similar calculations hold for all models; thus we always maintain a probability distribution as choice probabilities update throughout trials.
For the partial information setting, a linear model that converges to a probability matching strategy, as we prove in Appendix~\ref{sec:appendixA}, is the following model:
\\
For $i\in\{1,2,...,K\}$ and $0 \leq \alpha \leq 1$:\\ 
$ p_{i,n+1}=\begin{cases}
p_{i,n}+\alpha\cdot(1-p_{i,n}) & if\,\, j(n)=i\,\, and\,\, X_{i,n}=1\\
p_{i,n}-\alpha\cdot p_{i,n} & if\,\, j(n)=i\,\, and\,\, X_{i,n}=0\\
p_{i,n}-\alpha\cdot p_{i,n} & if\,\, j(n)\neq i\,\, and\,\, X_{j(n),n}=1\\
p_{i,n}+\alpha\cdot(\frac{1}{K-1}-p_{i,n}) & if\,\, j(n)\neq i\,\, and\,\, X_{j(n),n}=0
\end{cases} $
\\
Notice that for K=2 (i.e. the experimental task of the current study) the two models are identical, thus the full/partial information scenario result in the same update rule in the learning process. In other words, when people are facing only two shapes, a failure to receive a reward after selecting one shape is interpreted as if a reward was provided to the non-selected shape. We provide convergence proofs and fixed point analysis for both scenarios in Appendix~\ref{sec:appendixA}. In brief, the following model converges to a probability-matching strategy:
\begin{algorithm}[H]
\caption{Probability Matching Model}
Init: For $ i=1,2$ : $ p_{i,0} = \frac{1}{2}$ \\
for $ i=1,2 $ and $ 0 \leq \alpha \leq 1:$ \\
$ p_{i,n+1}=\begin{cases}
p_{i,n}+\alpha\cdot(1-p_{i,n}) & if\,\, X_{i,n}=1\\
p_{i,n}-\alpha\cdot p_{i,n} & otherwise
\end{cases}
$
\end{algorithm}

\paragraph{Modeling a maximization strategy}
Similar to the probability-matching model, the maximization model is based on a linear reinforcement learning update rule. The difference between the probability-matching and the maximization model stems from how the weights of the different shapes (or action choices) are updated in the event of negative outcomes. Specifically, in the maximization model (in contrast to the probability-matching model), the choice probabilities are not updated following a negative outcome and therefore remains identical to the previous trial. This model, as we prove analytically in Appendix~\ref{sec:appendixA}, converges to a maximization strategy:
\begin{algorithm}[H]
\caption{Maximization Model}
Init: For $ i=1,2,...,K $ : $ p_{i,0} = \frac{1}{K}$ \\
For $ i \in \{1,2,...,K\} $ and $ 0 \leq \alpha \leq 1 $ :\\
$p_{i,n+1}=\begin{cases}
p_{i,n}+\alpha\cdot(1-p_{i,n}) & if\,\, j(n)=i\,\, and\,\, X_{i,n}=1\\
p_{i,n} & if\,\, j(n)=i\,\, and\,\, X_{i,n}=0\\
p_{i,n}-\alpha\cdot p_{i,n} & if\,\, j(n)\neq i\,\, and\,\, X_{j(n),n}=1\\
p_{i,n} & if\,\, j(n)\neq i\,\, and\,\, X_{j(n),n}=0
\end{cases}
$
\end{algorithm}

\paragraph{Modeling the range of strategies between maximization and probability-matching}
This combined model is an interpolation between the probability update rules of the maximization and probability-matching models that allows any intermediate strategy between probability-matching and maximization. A full and formal proof can be found in Appendix~\ref{sec:appendixA}. For the case of K=2, the resulting model is:
\begin{algorithm}[H]
\caption{Combined Model}
Input: $t \in [0,1]$ \\
Init: For $ i=1,2 $ : $ p_{i,0} = \frac{1}{2}$ \\
For $i\in\{1,2\}$ and $0 \leq \alpha \leq 1$:\\ 
$ p_{i,n+1}=\begin{cases}
p_{i,n}+\alpha\cdot(1-p_{i,n}) & if\,\, j(n)=i\,\, and\,\, X_{i,n}=1\\
p_{i,n}-(1-\lambda)\cdot\alpha\cdot p_{i,n} & if\,\, j(n)=i\,\, and\,\, X_{i,n}=0\\
p_{i,n}-\alpha\cdot p_{i,n} & if\,\, j(n)\neq i\,\, and\,\, X_{j(n),n}=1\\
p_{i,n}+(1-\lambda)\cdot\alpha\cdot(1-p_{i,n}) & if\,\, j(n)\neq i\,\, and\,\, X_{j(n),n}=0 
\end{cases} $
\end{algorithm}
$\lambda$ is the interpolation factor that determines to what extent the learning strategy relates to maximization and probability-matching. Specifically, $\lambda = 0$ is equivalent to the probability-matching model, while  $\lambda=1$ is equivalent to the maximization model.

\subsection{Fitting the Model}
To provide inter-individual estimates of the learning strategies in each condition, we fitted the combined model for each participant in each condition. This procedure allows us to make inferences concerning the impact of different outcome Types and outcome Probabilities on the strategies applied under different conditions. The model fitting was achieved by estimating the maximum-likelihood parameters $\lambda, \alpha$. We emphasize here that we intentionally fit these parameters per condition, rather than fitting the model across the entire task, because our goal was to identify the correlations and causations between different outcome Types and Probabilities to the fitted parameters (and thus also the applied learning strategy). In other words, fitting the model per condition allows us to i) understand how the properties of the environment affect the behavioral policy across the population, and ii) ask questions regarding the variability of such policies within individuals.

\subsection{Statistical Procedures}
Learning performance and fitted model parameters were compared using repeated-measures ANOVAs, t-tests, and Pearson correlations. Corrections for multiple comparisons have been applied where needed via the Holm-Bonferroni method \citep{holm1979simple}. Learning performance is computed as the proportion of correct choice, i.e. the proportion of choosing the shape with the highest probability of receiving a positive outcome in each condition. 
K-means clustering over the nine-dimensional space of the fitted interpolation parameter (one $\lambda$ per condition and participant) was applied in an attempt to cluster participants into groups, which can then be compared in terms of behavioral parameters and fMRI data. By optimizing the mean silhouette value, which is a measure of how similar a point in a cluster is to points within its own cluster when compared to points in other clusters \citep{rousseeuw1987silhouettes}, the final number of clusters was determined to be K=2.
In a complementary analysis, Principal Component Analysis (PCA) was applied to the nine-dimensional space of the fitted interpolation parameter $\lambda$ in an attempt to find a subspace in which participants are well separated, and in which the main direction of variation can be extracted using the first PC. 

\subsection{MRI Data}

\paragraph{Image Acquisition}
MRI images were acquired using a 3T whole body MRI scanner (Trio TIM, Siemens, Germany) with a 12-channel head coil. Standard structural images were acquired with a T1 weighted 3D sequence (MPRAGE, Repetition time (TR)/Inversion delay time (TI)/Echo time (TE)=2300/900/2.98 ms, flip angle=9 degrees, voxel dimensions=1.0 mm isotropic, matrix size=256x256x176). Functional images were acquired with a susceptibility weighted EPI sequence (TR/TE=2000/27 ms, flip angle=75 degrees, voxel dimensions=3.0 mm isotropic, matrix size=187x153x118).

\paragraph{Data Analysis}
Functional MRI data were pre-processed and then analyzed using the general linear model (GLM) for event-related designs in SPM8 (Welcome Department of Imaging Neuroscience, London, UK; http://www.fil.ion.ucl.ac.uk/spm). During preprocessing, all functional volumes were realigned to the mean image, co-registered to the structural T1 image, corrected for slice timing, normalized to the MNI EPI-template, and smoothed using an 8 mm FWHM Gaussian kernel. Statistical analyses were performed on a voxel wise basis across the whole-brain. At the first-level, individual events were modeled by a standard synthetic hemodynamic response function (HRF). Twenty-four nuisance covariates (six rigid-body realignment parameters, their squares, their time derivatives, and their squared time derivatives) were also included when estimating statistical maps \citep{friston1996movement}. 

\paragraph{Model-based fMRI analysis}
To investigate the neural correlates of prediction errors (PEs), we used an event-related design which included three event-types time-locked to the onset of stimuli and positive and negative outcomes. Model-derived positive and negative PEs, calculated for each session separately, were added as linear parametric modulators to respective outcome type.

\paragraph{Region-Of-Interest (ROI) analysis}
Ventral tegmental area (VTA) ROI: PEs are coded by dopamine neurons in the ventral tegmental area (VTA; \citep{schultz2016dopamine}), and related findings have been reported using fMRI \citep{d2008bold}. Moreover, the differential encoding of positive versus negative PEs in the midbrain was related to learning biases \citep{aberg2015hemispheric}. Here, to test for neuronal correlates of positive and negative PEs and how they related to Maximization versus Probability-matching learning strategies, a VTA ROI was obtained from the Harvard Ascending Arousal Network Atlas \citep{edlow2012neuroanatomic}.

\paragraph{Statistical approach}
For each participant, the average beta parameter estimates across sessions for all voxels within the VTA ROI were extracted for each of the two PE types (Positive/Negative). Two different analyses were conducted to determine how the differential expression of learning strategies related to the neuronal coding of positive and negative PEs within the VTA ROI. In a first approach, a repeated measure ANOVA was performed with within-subjects factor PE-type (Positive, Negative) and Group (Probability-matching, Maximizing) as between-subject factor. Participants were classified according to the K-means clustering algorithm described previously. Paired and two-sample t-tests were used for follow-up analyses. In a second approach, the same repeated measures ANOVA was performed, except that the binary Group factor was replaced with a continuous covariate consisting of the first PC obtained from the PCA analysis described above. Pearson correlations were performed for follow-up analyses.

\paragraph{Whole-brain analysis}
For completion, we also analyzed the coding of positive and negative PEs across the whole-brain via family-wise error corrected (FWER) p-values. These analyses were conducted via t-tests implemented in SPM. An uncorrected initial search threshold of  p=0.0001 was used.
\section{Results}
\subsection{Behavior}
We start by testing how the three different outcome Types (-1, +1, and +3) and the three different outcome Probabilities (0.65, 0.75, 0.85) influence the probability of making a correct choice. A correct choice was defined as selecting the ‘best’ shape in each pair, i.e. the shape with the highest probability of obtaining a positive outcome (+1,+3; Conditions 1-6) or avoiding a negative outcome (-1; Conditions 7-9). The mean proportion of correct choices for the nine conditions as a function of trial are shown in Figure~\ref{fig:probMatchFig1}B, where the proportion of correct choice was calculated in a sliding window of 15 trials. Figure~\ref{fig:probMatchFig1}C shows the mean proportion of correct choice across the last 25 trials in each of the conditions. A two-way repeated measure ANOVA on the average proportion of correct choices in the last 25 trials with factors outcome Type (-1, +1, +3) and outcome Probability (0.65, 0.75, 0.85) revealed significant main effects of outcome Type [Figure~\ref{fig:probMatchFig1}D, lower panel; $F(2, 140)=6.797, p=0.0015$] and outcome Probability [Figure~\ref{fig:probMatchFig1}D, upper panel; $F(2, 140)=102.66, p<0.0001$], as well as a significant outcome Type x outcome Probability interaction [$F(4, 280)=4.9546, p<0.001$]. ]. To demonstrate robustness, we obtained the same effects when using both a smaller (last 15 trials) and larger (last 30 trials) window size. To elucidate these effects, a post-hoc analysis using the Holm-Bonferroni method was performed, with all reported p-values being the corrected p-values. First, the main effect of outcome Type was due to better overall performance for +1 outcomes [mean$\pm$SEM: $0.868\pm0.009$], as compared to +3 outcomes [mean$\pm$SEM: $0.822 \pm 0.01$; $t(70)=3.769, p=0.001$, $95\% \mbox{ } CI: 0.021 0.07$]. There were no significant differences between +1 and -1 outcomes, nor between +3 and -1 outcomes [all corrected p-values$>0.05$]. The main effect of outcome Probability was due to better overall performance for P=0.85 [mean$\pm$SEM: $0.925 \pm 0.007$] as compared to P=0.75 [mean$\pm$SEM: $0.877 \pm 0.011$, $t(70)=4.767, p<0.001, 95\% \mbox{ } CI: 0.028 0.069$], and P=0.65 [mean$\pm$SEM:$0.737 \pm 0.013$, $t(70)=13.995, p<0.001, 95\% \mbox{ } CI: 0.1616 0.215$], as well as better performance for P=0.75 versus P=0.65 [$t(70)=8.447,p<0.001, 95\% \mbox{ } CI: 0.106 0.172$]. For significant interaction effects, please see Table~\ref{table:t2}.
\begin{table}[H]
\begin{center}
 \begin{tabular}{||c c c c c c||} 
 \hline
 Interaction term & df & t-statistic & Corrected p-value & $95\%$ CI: Low & $95\%$ CI: High \\ [0.5ex] 
\hline\hline
0.85 vs 0.75 for +1 vs +3 & 70 & -2.535 & 0.021 & -0.092 & -0.007 \\
\hline
0.85 vs 0.65 for +1 vs +3 & 70 & -4.205	& $<0.001$ & -0.189 & -0.066 \\
\hline
0.85 vs 0.65 for -1 vs +3 &	70 & -2.997 & 0.004 & -0.19 & -0.038 \\
\hline
0.75 vs 0.65 for +1 vs +3 & 70 & -2.51 & 0.014 & -0.137 & -0.015 \\
\hline
\end{tabular}
  \caption{Significant interaction effects for outcome Probability x outcome Type. All presented p-values are calculated using the Holm-Bonferroni method. Only interaction terms with corrected p-values $< 0.05$ are displayed.}
\label{table:t2}
\end{center}
\end{table}

Our results indicate that both factors as well as their interaction affect the variance in learning performance. To determine whether participants had converged in their choice preference (i.e. whether learning was still on-going at the end of each session), we linearly regressed the proportion of correct choices for the first ten (trial 1-10) and for the last ten (trial 41-50) trials for each individual and condition separately. We then performed a three-way repeated measure ANOVA on the slopes, with factors outcome Type (-1, +1, +3), outcome Probability (0.65, 0.75, 0.85), and Slope timing (first ten trials, last ten trials). This analysis revealed significant effects of outcome Probability [$F(2, 140)= 7.426, p<0.001$], Slope timing [$F(1, 70)= 51.97, p<0.0001$], and their interaction [$F(2,140) = 7.7175, p<0.001$]], as well as the interaction between Outcome Type and outcome Probability [$F(4,280) = 2.636, p<0.05$] and outcome Type x outcome Probability x Slope timing interaction [$F(4,280) = 2.874, p<0.05$].  No other effects or interactions were significant (all p-values$>0.05$). As would be expected, more learning occurred initially, as indicated by larger mean values for the slopes of the first ten trials [mean$\pm$SEM: $0.015 \pm 0.001$] as compared to the last ten trials [mean$\pm$SEM: $0.001 \pm 0.0007$]. The mean slopes of the first and last 10 trials, as well as the average slopes for trials 10-19 and 20-29 are shown in Figure~\ref{probMatchFig1}E. Importantly, a two-way repeated measures ANOVA with the same factors as above (i.e. outcome Type and Probability) and slopes for the last ten trials as dependent variable revealed a non-significant intercept term [$F(1, 70)=2.39, p=0.127$]. This result, indicating that the mean slope value across conditions for the last ten trials is not significantly different from 0.0, confirms that learning had converged. No other effects or interactions were significant (all p-values $> 0.05$). A similar ANOVA for the initial ten trials revealed an intercept term significantly different from 0.0 [$F(1,70)=73.078, p<0.0001$], as well as a significant main effect of outcome Probability [$F(2,140)=8.869, p<0.001$]. This latter result is because the initial learning slopes were larger for P=0.85 [mean$\pm$SEM:$0.025 \pm 0.003$], as compared to P=0.75 [mean$\pm$SEM:$0.012 \pm 0.003$; $t(70)=2.823, p=0.012, 95\% \mbox{ } CI: 0.004 0.022$] and as compared to P=0.65 [mean$\pm$SEM:$0.008 \pm 0.002$; $t(70)=3.886, p<0.001, 95\% \mbox{ } CI: 0.008 0.026$]. Outcome Type x outcome Probability interaction was also significant [$F(4,280)=2.939, p-value<0.05$]. No other effects or interactions were significant (all p-values $> 0.05$). \\

In summary, our result show that participants successfully learned the task. Not surprisingly, performance increased with increasing outcome Probability (i.e. with decreasing difficulty). Slightly surprising however, was that performance did not scale monotonically with outcome Type (i.e. the highest performance was reached for R=+1 and not for R=+3). Yet, non-monotonic relationships between reward feedback and learning has been observed previously \citep{aberg2020interplay}. Next, we addressed the mechanisms underpinning an individuals' tendency to engage in maximization versus probability-matching strategies.
\subsection{Computational Modeling} 
\paragraph{Model simulations of behavior} \mbox{ } \\
In-line with recent recommendations on computational modeling of behavior \citep{wilson2019ten}, we start by showing the power of our proposed model through simulations. All simulations were run in a setting with a triplet of shapes (outcome Probabilities 0.10, 0.25, and 0.65), outcome Type set to +1, and learning rate set to 0.01. Please observe that the first two shapes are the ‘worst’ shapes and should be avoided, while the third shape is the ‘best’ shape and should be chosen. These simulations were run using a triplet of shapes for illustrative purposes, and simulations on pairs of shapes are provided below. \\

\begin{figure}[H]
\centering
\includegraphics[scale=0.8]{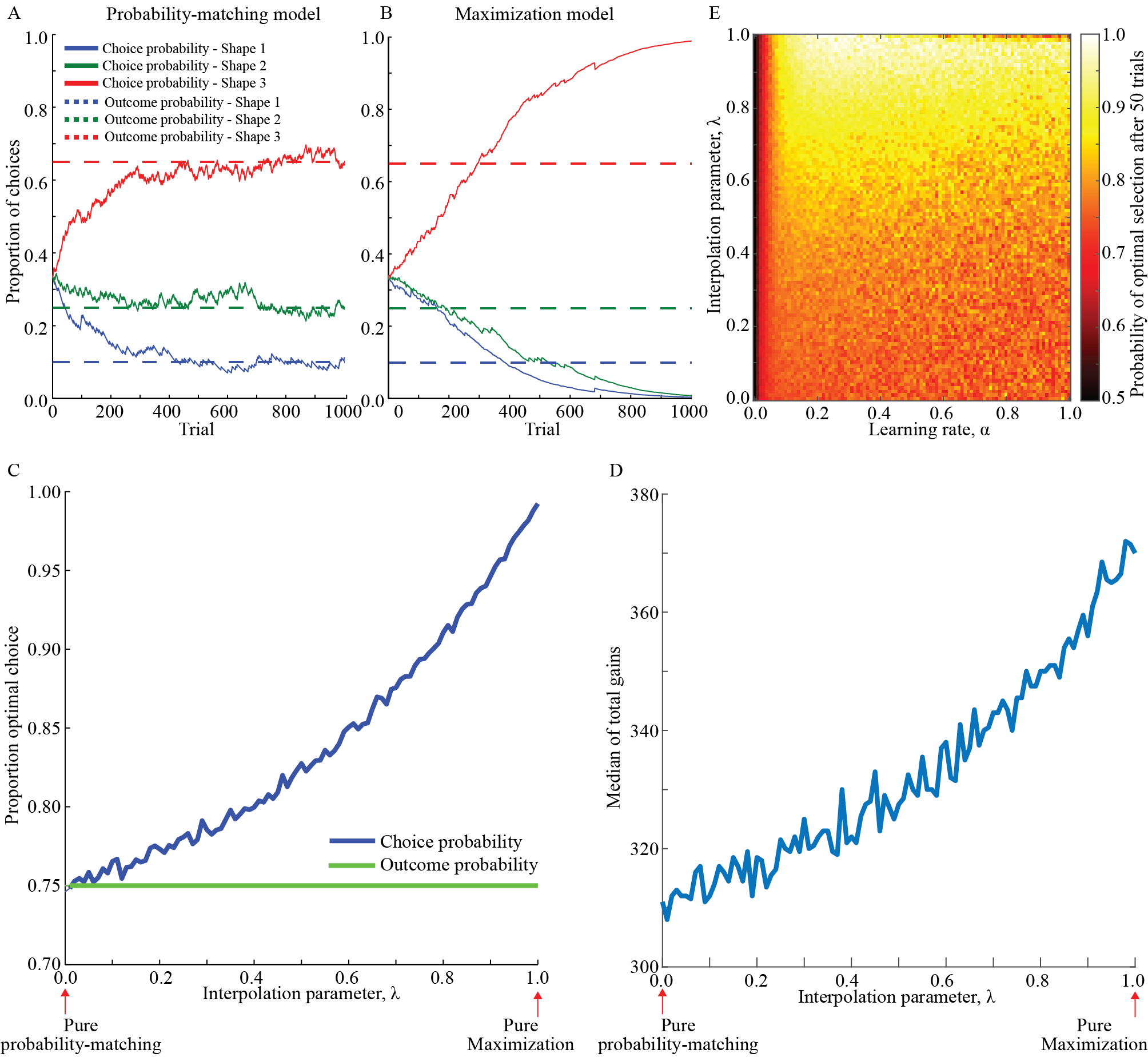}
\caption{Model-simulated performance. For the probability-matching model (A), the probability of selecting each option is equal to its outcome probability, while for the maximization model (B) the probability of selecting the ‘best’ and ‘worst’ options is equal to 1 and 0 respectively. All simulations were run with 1000 trials, outcome Type = +1, and learning rate $\alpha = 0.01$. The dashed lines represent the outcome Probabilities of the three different shapes. The solid lines represent the model-derived probability of choosing each shape. C. The model-derived probability of choosing the ‘best’ shape in a pair as a function of interpolation parameter $\lambda$ for the combined model. Each point in the graph represents the convergent probability to choose the ‘best’ shape. All simulations were run with 1000 trials and learning rate $\alpha = 0.01$. The outcome Probability and Type were set to P=0.75 and R=+1. D. The median total gains as a function of the interpolation parameter $\lambda$. E. The model-derived probability of choosing the ‘best’ shape after 50 trials as a function of different learning rates $\alpha$ and interpolation parameters $\lambda$. The outcome Probability and Type were set to P=0.75 and R=+1.}
\label{fig:probMatchFig2}
\end{figure}

As expected according to the model convergence analysis (see Appendix~\ref{sec:appendixA}), behavior converges to probability-matching (i.e. the proportion of choices for a particular shape corresponds to its outcome probability) for the probability-matching model (Figure~\ref{fig:probMatchFig2}A) and to maximization (the proportion of choices for the ‘best’ shape converges to 1.0 while the proportion of choosing any other shape converges to 0.0), for the maximization model (Figure~\ref{fig:probMatchFig2}B).
To illustrate the impact of the combined model, we simulated the probability of selecting the ‘best’ shape in a pair of shapes with outcome Probability 0.85 as a function of the interpolation parameter $\lambda$ after 1000 trials (Figure~\ref{fig:probMatchFig2}C). As expected, low values of $\lambda$ lead to more probability-matching while increasing values of $\lambda$ lead to more maximization. Moreover, to illustrate that probability-matching is suboptimal in the present task with static outcome Probabilities, we simulated the accumulated reward for different values of $\lambda$. Indeed, total reward was lowest for pure probability matching ($\lambda=0$) and highest for pure maximization ($\lambda=1$; Figure~\ref{fig:probMatchFig2}D). 

To be more close to our experimental data, where 50 trials were provided in each condition, we also simulated the combined model with 50 trials for different values of the interpolation parameter $\lambda$ and the learning rate $\alpha$. The probability of selecting the ‘best’ shape in a pair of shapes (here, with an outcome Probability 0.75) after 50 trials is shown in Figure~\ref{fig:probMatchFig2}E. Figure~\ref{fig:probMatchFig2}E demonstrates that for low values of $\lambda$, and $\alpha > 0.05$, the choice probability converges to probability matching, while maximization becomes more prominent for larger values of $\lambda$. For very low values of $\alpha$ ($<0.05$), the value of $\lambda$ does not affect the convergent choice probability because more trials are needed to reach convergence. For this reason, six participants with a fitted learning rate lower than $0.05$ in a majority of conditions were excluded from the analyses. We note that including these six participants does not change the results or the conclusions. \\

In summary, these simulations show that the combined model is capable of capturing the range of behaviors from probability-matching to maximization and that the interpolation parameter $\lambda$ determines the convergent choice probability. \\

\begin{figure}[H]
\centering
\includegraphics[scale=0.8]{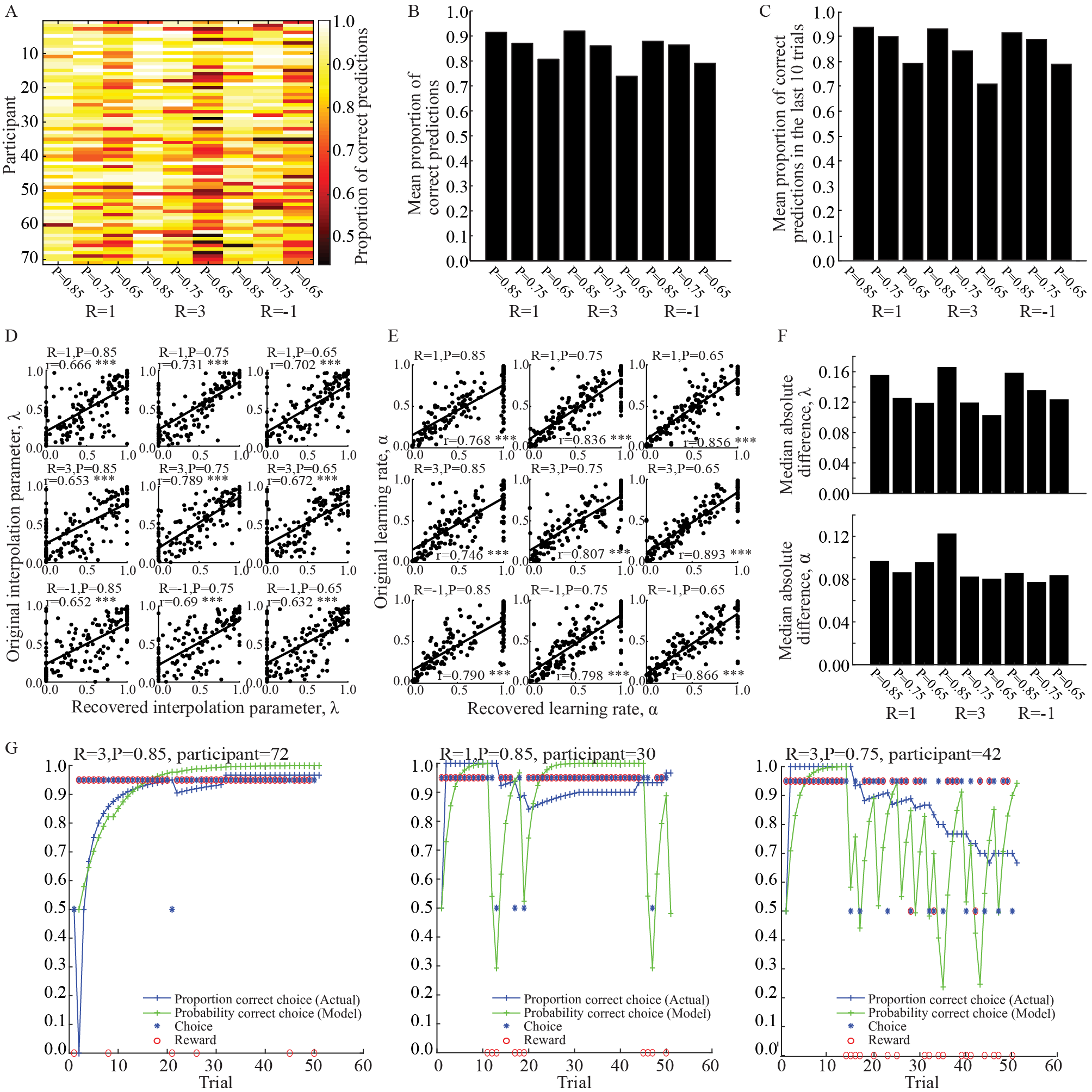}
\caption{Validation of the combined model. A. Proportion of model-derived correct predictions per subject and condition. The proportion of correct predictions is computed as the proportion of trials in which the model predicts the same choice as the participant. B. The mean proportion of model-derived correct prediction in all conditions. C. The mean proportion of model-derived correct predictions in the last 10 trials when using the model parameters fitted for the 40 first trials. The error bars indicate SEM. D. Correlations between actual and recovered interpolation parameters $\lambda$ in all 9 conditions for 150 virtual participants (all Holm-Bonferroni corrected p-values $< 0.0001$). E. Correlations between actual and recovered learning rates $\alpha$ in all 9 conditions for 150 virtual participants (all Holm-Bonferroni corrected p-values $< 0.0001$). F. Median difference between actual and recovered interpolation parameters $\lambda$ (top) and learning rate $\alpha$ (bottom). G. Individual representative examples of model fits to behavior. Blue line: the actual proportion of correct choice calculated in a window of 20 trials back. Green line: the model-derived probability to choose the correct option. Blue asterisks (*): participant choices, where a blue * located above y=0.5 indicates a correct choice, whereas the worst shape was chosen otherwise (i.e. when a blue * is located at y=0.5). Red circles: outcomes, where positive outcomes are denoted by a circle overlapping the choice, while circles located at y=0 denote negative outcomes 0. The leftmost panel shows a participant that is relatively insensitive to negative outcomes, as indicated by few changes in behavior following negative outcomes, and this maximization behavior is captured by the computational model. By contrast, the middle and rightmost panels show participants that are more sensitive to negative outcomes, and the correspondent increases in probability-matching behaviors are also captured by the model.}
\label{fig:probMatchFig3}
\end{figure}

\paragraph{Fitting the combined model to behavioral data } \mbox{ } \\
The model simulations suggest that the combined model is capable of capturing relevant aspects of learning performance, yet it is critical to assess how well the model fits actual behavior. As a first validation, the combined model was fitted to behavioral data for each participant and condition separately (i.e. for each participant nine different values of the interpolation parameter $\lambda$ and the learning rate $\alpha$ was obtained). The fitted parameters were then used to obtain model-derived predictions for which shape would be selected in each trial. If the model assigns a higher probability for a shape that was actually selected, a score of 1 was awarded while a 0 is provided otherwise. Figure~\ref{fig:probMatchFig3}A shows the proportion of correct predictions made by the combined model for each participant in each condition, and Figure~\ref{fig:probMatchFig3}B shows the mean value of the proportion correct predictions in each condition. The model achieves high accuracy in all conditions. As a second validation, the interpolation parameter $\lambda$ and the learning rate $\alpha$ were fitted using only the first 40 trials, and the accuracy of the model in the last 10 trials was then assessed (using the same procedure described above). Figure~\ref{fig:probMatchFig3}C shows the mean percent of correct predictions in each condition. Similar to fitting the parameters across all trials, the model achieves high accuracy in all conditions. As a third and final validation, 150 virtual participants were created by randomly sampling different interpolation parameters $\lambda$ and learning rates $\alpha$ (both ranging from 0 to 1), and then simulate their performance in each condition. Next, we recovered each virtual participants' individual parameters by fitting the combined model to their simulated performance data. Figures~\ref{fig:probMatchFig3}D and ~\ref{fig:probMatchFig3}E show the correlation between the original versus the recovered interpolation parameter $\lambda$ and learning rate $\alpha$, correspondingly, in each of the nine conditions. All correlations are highly significant, with all Holm-Bonferroni corrected p-values $< 0.0001$. In an alternative display of these results, the median absolute difference between the actual and the recovered interpolation parameters $\lambda$ and learning rates $\alpha$ are respectively shown in Figure~\ref{fig:probMatchFig3}F, top and bottom panels. Importantly, the actual values of $\lambda$ and $\alpha$ are both in the range of  [0,1], whereas the mean differences shown in Figure~\ref{fig:probMatchFig3}F (right panel) are an order of magnitude lower than that, again indicating a small difference between actual and recovered parameters. Accordingly, because the actual and recovered parameters are very similar they can be considered as being meaningful \citep{wilson2019ten}.
For visualization purposes, Figure~\ref{fig:probMatchFig3}G shows three examples of representable participants and their fits under different conditions. These examples illustrate that the model is able to correctly predict the participant's choices and capture the learning process in these different scenarios. \\

\begin{figure}[H]
\centering
\includegraphics[scale=0.8]{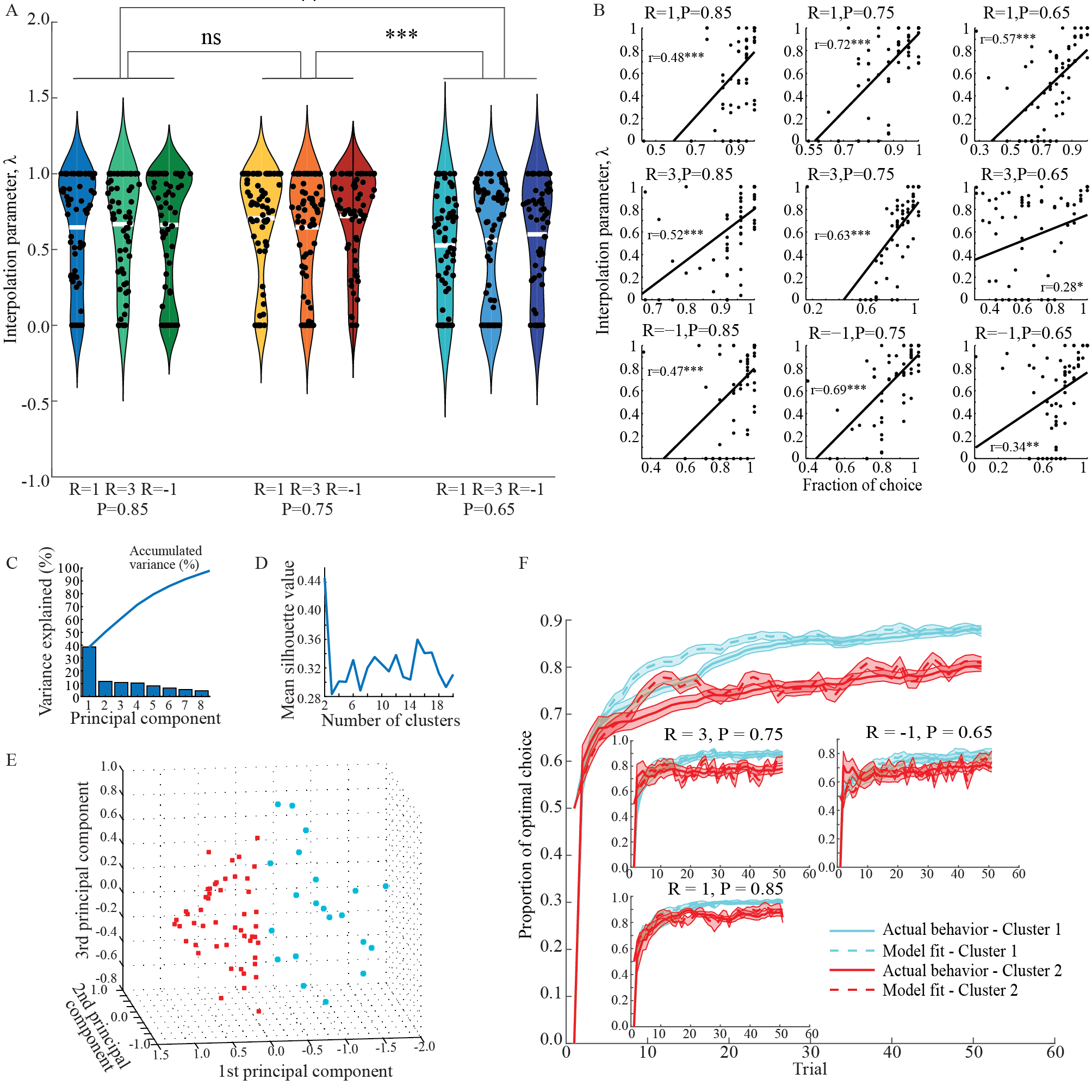}
\caption{Learning strategies across conditions and within individuals. A. The distribution of the interpolation parameter $\lambda$ per condition. P and R respectively indicates outcome Probability and Type. B. Correlations between the interpolation parameters $\lambda$ and the proportion of correct choice in each condition. All correlations are positive and significant after applying Holm-Bonferroni correction. C. Percent variance explained for each principal component, and accumulated variance (solid line). D. Mean silhouette value for clusters ranging from K=2 up to K=20. K=2 maximizes the mean silhouette value. E. K-means (K=2) clustering of participants by the nine-dimensional vector of interpolation parameters $\lambda$. The clusters are clearly separated by the first (out of the three) principal component(s) obtained from a PCA analysis using the same nine-dimensional vector as input. F. To illustrate the behavior of participants in the two clusters, the mean proportion correct and the mean model-fitted proportion correct over all conditions are shown in the main figure, and for three individual conditions in the insets: P=0.85 and R=+1, P=0.75 and R=+3, P=0.65 and R=-1. The behavior of participants in Cluster 1 is reminiscent of a maximization strategy, while participants in Cluster 2 behave more similarly to a probability matching strategy. ns not significant, ***$p< 0.001$, **$p< 0.01$, *$p<0.05$.}
\label{fig:probMatchFig4}
\end{figure}

\paragraph{Learning strategy depends on outcome Probability, but not outcome Type} \mbox{ } \\
We tested whether the learning strategy differed between outcome Type and/or outcome Probability by testing for differences in the interpolation parameter $\lambda$ between conditions. Figure~\ref{fig:probMatchFig4}A shows the distribution of the $\lambda$’s for each condition. A repeated measures ANOVA with factors outcome Type (-1, +1, +3) and outcome Probability (0.65, 0.75, 0.85) revealed a significant main effect of outcome Probability [$F(2, 140) = 9.951, p<0.001$]. This effect was caused by larger values of $\lambda$ for p=0.75 [mean$\pm$SEM:$0.69 \pm 0.03$] compared to 0.65 [mean$\pm$SEM:$0.563 \pm 0.03$; $t(70)=4.683, p<0.001, 95\% \mbox{ } CI: 0.073 0.182$], and larger values of $\lambda$ for p=0.85 [mean$\pm$SEM:$0.658 \pm 0.033$] compared to p=0.65 [$t(70)=3.152, p=0.004, 95\% \mbox{ } CI: 0.034 0.155$]. By contrast, there was no difference in learning strategy between P=0.85 and P=0.75 [$t(70)=1.033, p=0.305, 95\% \mbox{ } CI: -0.095 0.03$]. All reported p-values were corrected using the Holm-Bonferroni method. Moreover, the main effect of outcome Type was not significant [$F(2, 140) = 1.094, p=0.337$] and neither was the outcome Type x Probability interaction [$F(4, 280) = 0.658, p=0.621$]. These results suggest that participants (in general) show a stronger tendency to apply a probability-matching strategy when the task is more difficult (i.e. with small outcome probabilities), while a maximization strategy is more likely when the task is easier.\\
\paragraph{Learning strategy correlates with performance} \mbox{ } \\
The model simulations show that the interpolation parameter $\lambda$ correlates positively with the proportion of correct choice (see Figure~\ref{fig:probMatchFig2}C). To confirm such a relationship also in actual performance data, we correlated the model-fitted $\lambda$ parameter in each condition with the corresponding proportion of correct choices in the last 25 trials. As expected, positive and significant correlations were obtained in each of the nine conditions (all Pearson’s $r>0.28$; all Holm-Bonferroni corrected p-values$<0.05$; see Figure~\ref{fig:probMatchFig4}B for individual correlations in each condition). A significant correlation (Pearson’s $r=0.471$, $p-value<0.0001$) was also found between the model-fitted $\lambda$ parameter and proportion of correct choices in the last 25 trials when correlating across all conditions. \\

\paragraph{Learning strategy is consistent between conditions and within an individual} \mbox{ } \\
To test whether individual participants expressed consistent learning strategies across conditions, we calculated Cronbach’s alpha coefficient for the vector of nine $\lambda$’s. In brief, Cronbach’s alpha coefficient estimates how well a test measures what it was designed to measure, and is frequently used to assess the reliability of multiple-item tests or questionnaires \citep{tavakol2011making}. For example, a questionnaire designed to measure trait anxiety, with a high enough Cronbach’s alpha coefficient, can be said to measure trait anxiety, while a low alpha coefficient suggests it does not (i.e. there is low internal consistency between items that are supposed to measure the same construct). In the context of the present study, we use Cronbach’s alpha to estimate whether participants show a consistent expression of learning strategies. Indeed, Cronbach’s alpha coefficient, when treating each $\lambda$ as an item, was 0.796, which is consistent with an acceptable/good internal consistency \citep{tavakol2011making}. In other words, these results suggest that a learning strategy can be regarded as a consistent and trait-like feature of an individual.

\begin{figure}[H]
\centering
\includegraphics[scale=1]{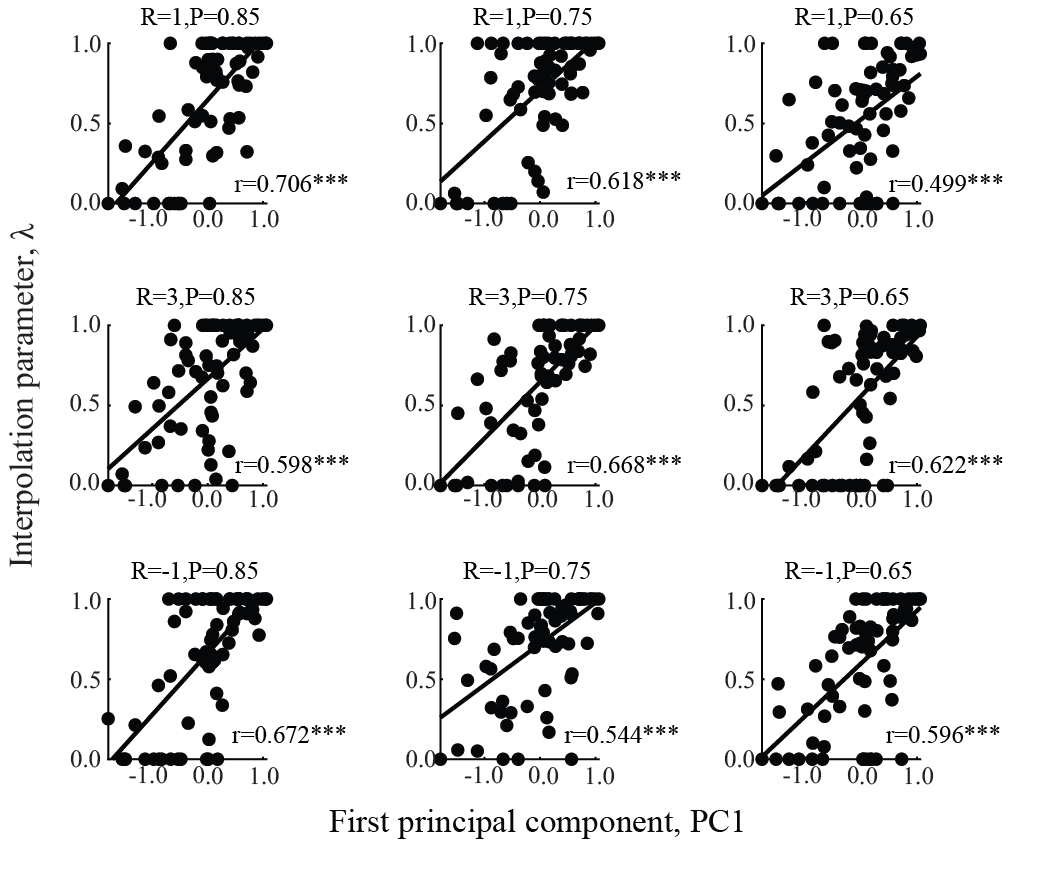}
\caption{Pearson’s correlations between the first principal component (PC1) and the interpolation parameter $\lambda$ in each condition. ***$p<0.001$.}
\label{suppFig:pm_supp1}
\end{figure}

To address this idea further, we condensed relevant variance across the nine $\lambda$’s into three dimensions via Principal Component Analysis (PCA; see Methods). The first PC captured $38.5\%$ of the total variance (Figure~\ref{fig:probMatchFig4}C). To test the relation between the first PC and the original dimensions (the $\lambda$’s of the nine different conditions), we correlated the first PC of each participant with each of their nine session-specific $\lambda$’s. As would be expected, the first PC correlated significantly with the $\lambda$’s in all nine conditions (all Holm-Bonferroni corrected p-values$<0.001$, Supplementary Figure~\ref{suppFig:pm_supp1}). Importantly, this result also shows that no single dimension uniquely defines the direction of the first PC. Together, these results suggest that the first PC indeed captures relevant variance in learning strategy. \\

\begin{figure}[H]
\centering
\includegraphics[scale=0.8]{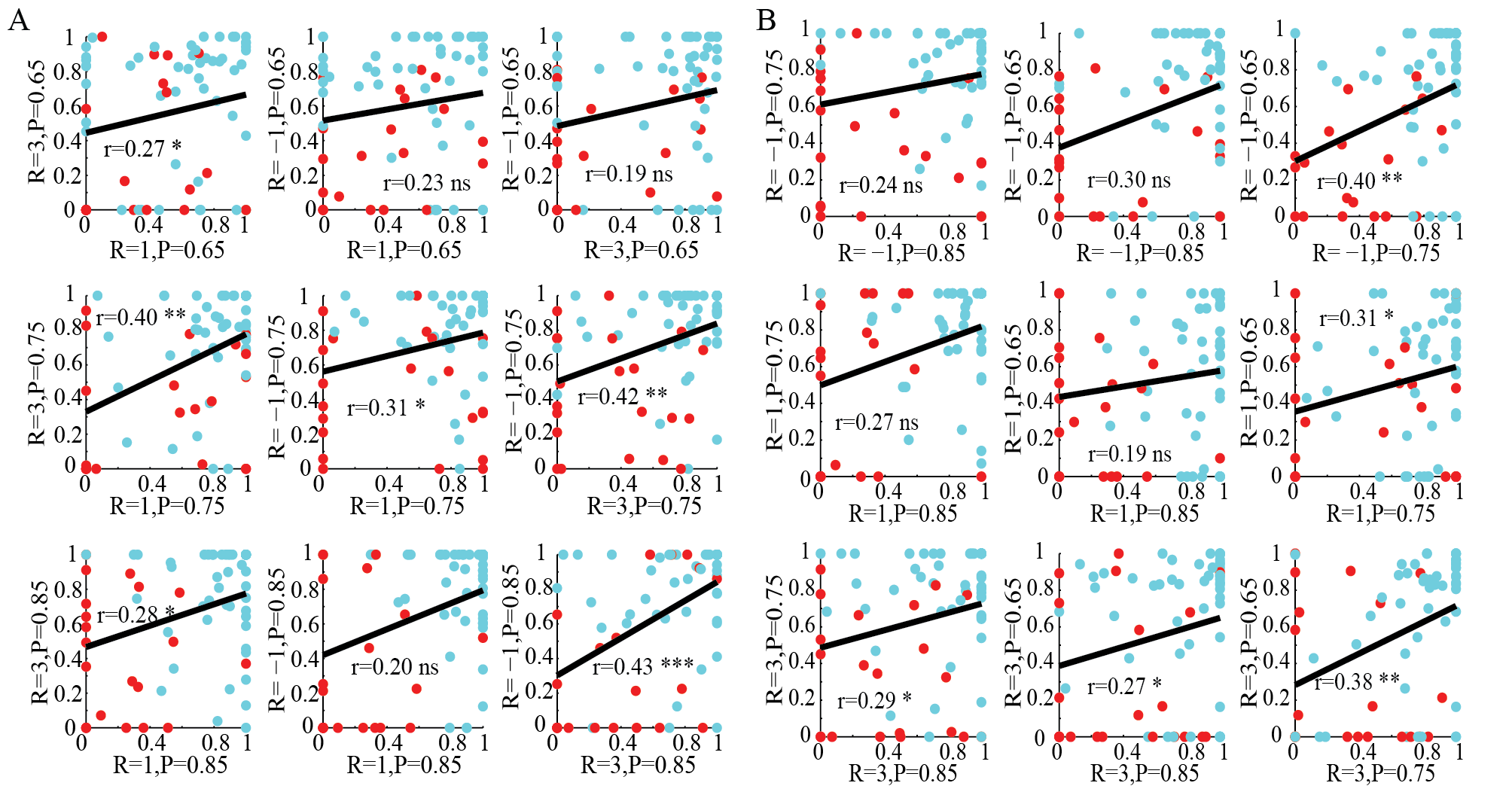}
\caption{Correlations between interpolation parameters $\lambda$ in different conditions. A. Correlation between interpolation parameters $\lambda$ in pairs with the same outcome Probability but different outcome Type. B. Correlation between interpolation parameters $\lambda$ in pairs with the same outcome Type but different outcome Probabilities. Colors (red/turquoise) correspond to the participant’s cluster assignment. All p-values are corrected using the Holm-Bonferroni method. ***$p<0.001$, **$p<0.01$, *$p<0.05$. ns not significant ($p>0.05$).}
\label{suppFig:pm_supp2}
\end{figure}

To further validate these results, we tested whether the first principal component (PC1) captured variance related to task performance, given the significant correlations between $\lambda$ and the proportion correct (Figure~\ref{fig:probMatchFig4}B). For this reason, a repeated measures ANOVA with factors outcome Type (-1, +1, +3) and outcome Probability (0.65, 0.75, 0.85) was conducted with PC1 as continuous covariate. This analysis revealed a significant main effect of PC1 [$F(1,69)=57.287, p<0.001$], but no interactions with PC1 (all p-values$>0.05$). To clarify, there was a positive correlation between PC1 and the average proportion correct collapsed across conditions [Pearson’s $r=0.673, p<0.0001$]. This result confirms that the PC1 captures variance in learning strategy that is relevant to explain inter-individual differences in performance across conditions. Finally, the interpolation parameters $\lambda$ were correlated between different pairs of conditions with different outcome Types but with the same outcome Probability. All of the nine possible correlations were positive with uncorrected p-values $\leq 0.10$, and six of the correlations survived Holm-Bonferroni correction (Supplementary Figure~\ref{suppFig:pm_supp2}A). In addition, correlating the interpolation parameters $\lambda$ between different pairs of conditions with different outcome Probabilities but with the same outcome Type revealed similar results, with all of the nine possible correlations being positive with uncorrected p-values $\leq 0.10$, and three of these survived Holm-Bonferroni correction (Supplementary Figure~\ref{suppFig:pm_supp2}B). Correlations between pairs with both different outcome Types and Probabilities were not conducted because participants tend to apply slightly different learning strategies for different outcome probabilities (see preceding paragraph; Figure~\ref{fig:probMatchFig4}A). \\

In summary, these results suggest that individuals express consistent and trait-like learning strategies across task conditions.\\

\paragraph{Learning strategy partitions learners into two groups} \mbox{ } \\
A K-means clustering algorithm was applied to test whether participants could be partitioned into different groups based on their learning strategy in the different conditions. Using the nine-dimensional vector of $\lambda$’s for each participant as input, this approach revealed that participants could be divided into two clusters (K=2). The number of clusters was chosen by optimizing the mean silhouette value, which is a measure of how similar a point in a cluster is to points within its own cluster as compared to points in other clusters (Figure~\ref{fig:probMatchFig4}D). \\

\begin{figure}[H]
\centering
\includegraphics[scale=0.8]{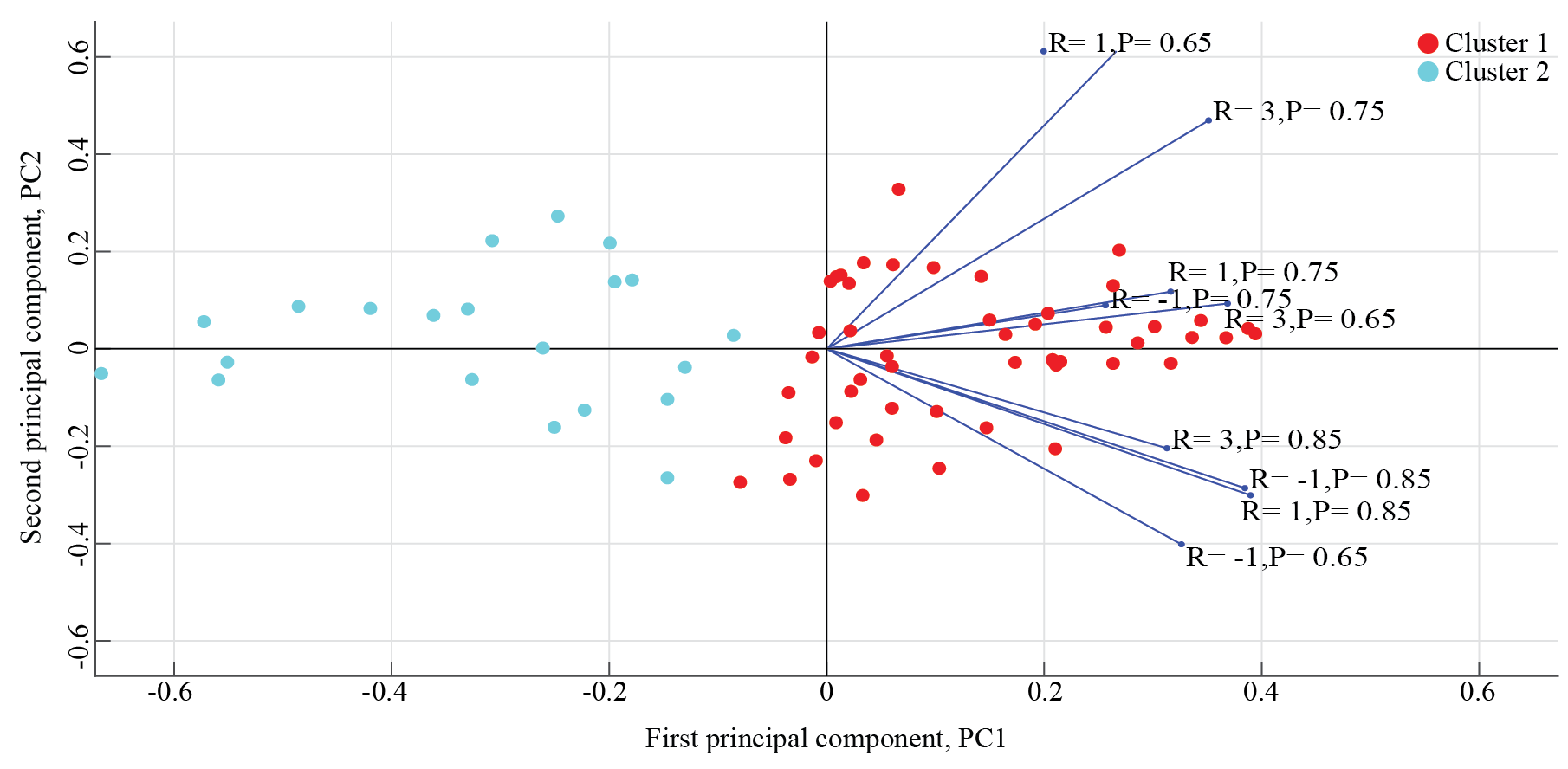}
\caption{Projection of the original nine dimensions (the nine conditions with different outcome Type and outcome Probability) on the first and second principal components (PC1, PC2). The results show i) that the first PC provides a good separation between the two clusters, and ii) that no single dimension determines the first PC. Dots represent the projection of each individual in the space of the first and second PC component, where the colors correspond to the cluster assignments. Red: Cluster 1 (probability matchers), Blue: Cluster 2 (maximizers).}
\label{suppFig:pm_supp3}
\end{figure}

To visualize the relationship between the PCA and the K-means clustering approach, we plot the projection of each of the nine $\lambda$’s in the space of the first and second PC and the projection of each individual, color-coded by the cluster to which they were categorized, in that space. The first PC uniquely separates between the clusters (Supplementary Figure~\ref{suppFig:pm_supp3}; Figure~\ref{fig:probMatchFig4}E displays the first three PCs). \\

To illustrate the behavior of participants in the two clusters, we plot the mean proportion of correct choice over all conditions (Figure~\ref{fig:probMatchFig4}F), and the mean proportion of correct choice in three individual conditions (Figure~\ref{fig:probMatchFig4}F, insets). Individual plots for all nine conditions are shown in Supplementary Figure~\ref{suppFig:pm_supp4}. The two clusters clearly converge to different choice probabilities, with participants in Cluster 1 pertaining more to a maximization strategy while participants in Cluster 2 show more probability-matching. Moreover, the model-derived choice probability, using the individually fitted parameters of each individual in each cluster, closely match the actual behavior of participants in respective clusters (Figure~\ref{fig:probMatchFig4}F). \\

For simplicity, from now on we refer to these two different clusters of participants as the ‘Maximizing’ group and the ‘Probability-matching’ group. 

\begin{figure}[H]
\centering
\includegraphics[scale=0.8]{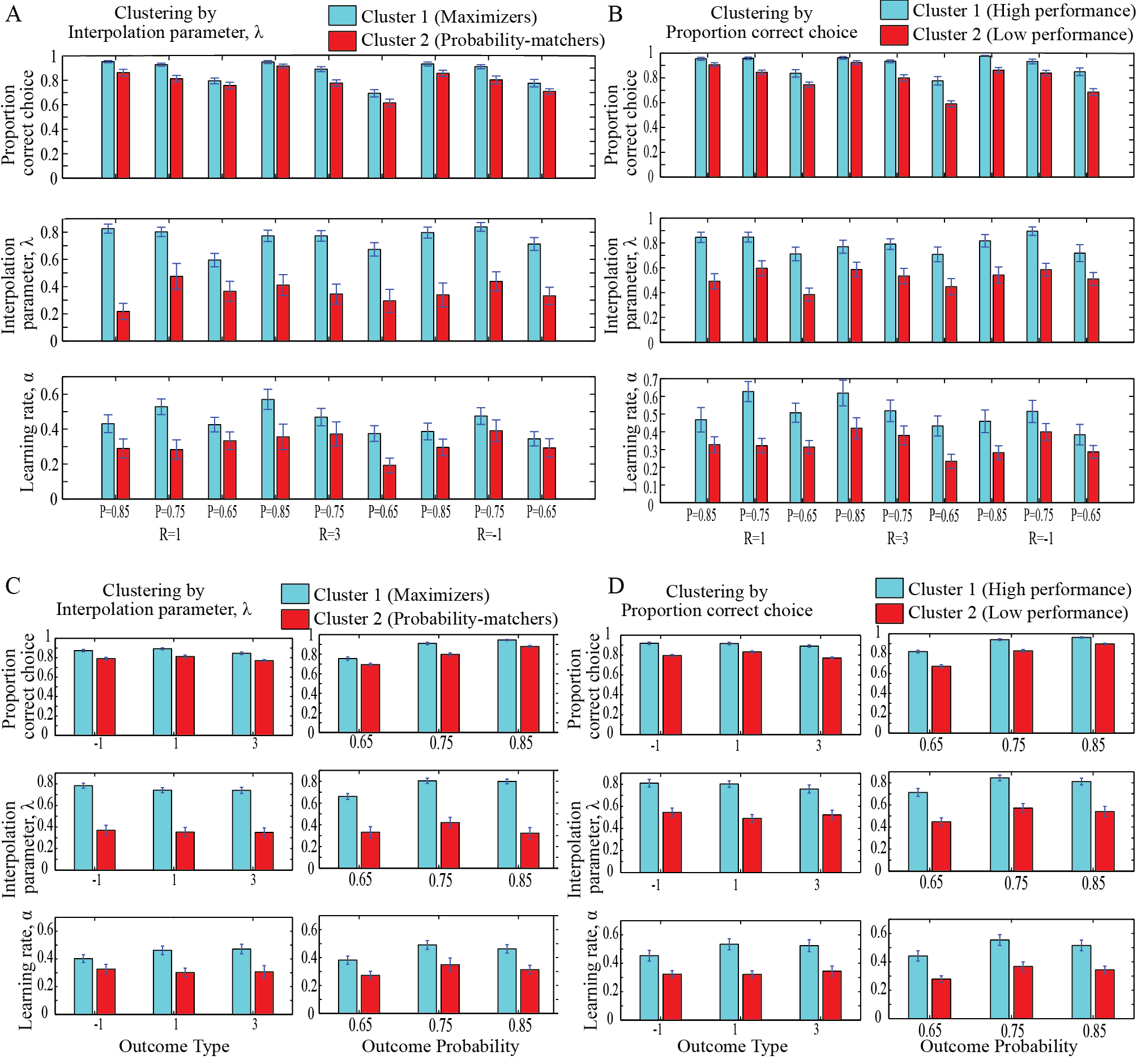}
\caption{Cluster characteristics. A. Mean characteristics of the two clusters defined by the 9-dimensional vector of the interpolation parameters $\lambda$. B. Mean characteristics of the two clusters defined by the 9-dimensional vector of the proportion of correct choices. C. Mean characteristics of the two clusters defined by the 9-dimensional vector of the interpolation parameters $\lambda$ collapsed across outcome Type (left columns) and outcome Probability (right columns). D. Mean characteristics of the two clusters defined by the 9-dimensional vector of the proportion of correct choices collapsed across outcome Type (left columns) and outcome Probability (right columns).}
\label{suppFig:pm_supp5}
\end{figure}

\paragraph{Characteristics of the Maximizing and Probability-matching groups} \mbox{ } \\
Three characteristics of the participants in the two groups are investigated further:  the mean proportion of correct choice, the mean interpolation parameter $\lambda$, and the mean learning rate $\alpha$. We observe that the Maximizing group show consistently higher values in each of the three characteristics across conditions (Supplementary Figure~\ref{suppFig:pm_supp5}A,C). Confirming this observation, a repeated measures ANOVA with Group as a between subjects factor, and outcome Probability and outcome Type as within subject factors, revealed a significant main effect of cluster assignment for all three characteristics [Proportion correct choice: $F(1,69)=32.43, p<0.0001$; Interpolation parameter $\lambda$: $F(1,69)=142.47, p<0.0001$; Learning rate $\alpha$: $F(1,69)=12.762, p=0.0006$]. No interactions with the Group factor were significant (all p-values$>0.05$). It is important to note that the findings that all characteristics in all conditions are consistently different between the two clusters are not trivial, because the clusters are globally defined and are not a property of any single condition. For example, the same clusters could have been obtained if the $\lambda$’s in some conditions had been inverted, i.e., where in some conditions the $\lambda$’s would be higher for cluster 1 than for cluster 2, and in other conditions lower for cluster 1 compared to cluster 2. Still, we observed that the characteristics of the clusters stay consistent throughout the conditions. In summary, these results support our previous results showing that the interpolation parameter and proportion of correct choice are closely correlated. \\

We also conducted a complementary analysis in which participants were clustered based on the proportion of correct choice (averaged across the last 25 trials) in each condition (i.e. rather than the interpolation parameter $\lambda$). Again, using the same optimization criterion (i.e. the silhouette value), two separate clusters of participants were defined and the same three characteristics were compared for the two different clusters (Supplementary Figure~\ref{suppFig:pm_supp5}B,D). Applying the same analyses as in the previous paragraph, we obtain the same results, namely a significant main effect of cluster assignment for all three characteristics [Proportion correct choice: $F(1,69)=159.6, p<0.0001$; Interpolation parameter $\lambda$: $F(1,69)=40.077, p<0.0001$; Learning rate $\alpha$: $F(1,69)=31.705, p<0.0001$]. 
\subsection{fMRI}
\label{sec:fmri}
\begin{figure}[H]
\centering
\includegraphics[scale=1]{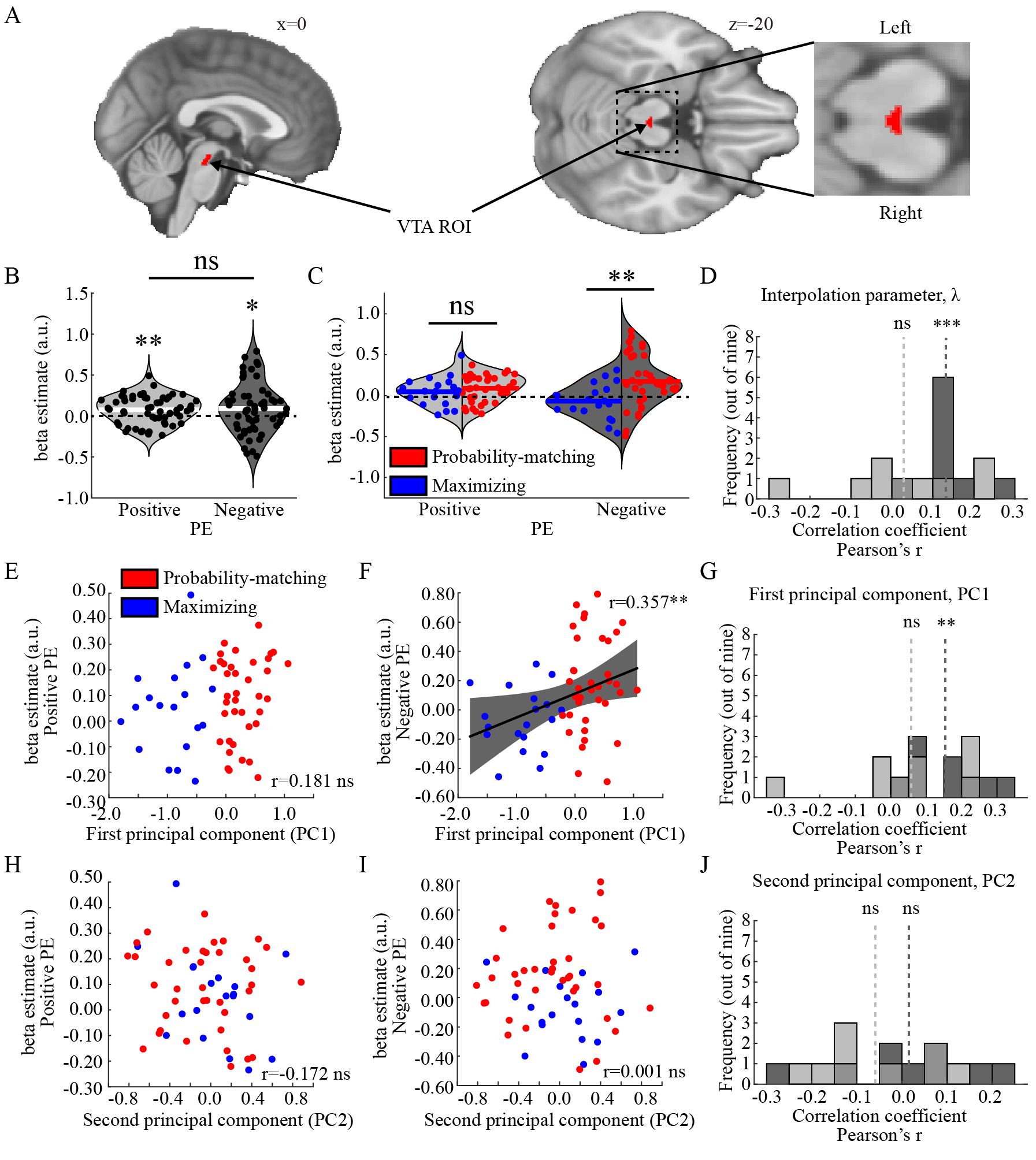}
\caption{Functional MRI (fMRI) correlates of prediction errors (PEs) in a ventral tegmental area (VTA) region of interest (ROI). A. The VTA ROI used for extracting beta parameter estimates for Positive and Negative PEs. B. Across participants, both PE-types correlate significantly with VTA ROI BOLD signal. C. Participants in the Probability-matching (vs. Maximizing) group show significantly stronger correlation between VTA BOLD signal and Negative PEs, while there was no significant group-difference for Positive PEs. D. The distribution of correlation coefficients (one Pearson’s r was calculated for each of the nine conditions) between condition-specific interpolation parameter $\lambda$ and the coupling between VTA BOLD signal and Positive PEs (light grey) and Negative PEs (dark grey). On average, this correlation is significant for Negative PEs but not for Positive PEs (average $r > 0$; two-tailed t-tests). E. The first PCA component (PC1) does not correlate significantly with the VTA representation of Positive PEs, but show a positive correlation for Negative PEs (F). G. The distribution of correlation coefficients (one Pearson’s r calculated for each of the nine conditions) between PC1 and the coupling between VTA BOLD signal and Positive PEs (light grey) and Negative PEs (dark grey). On average, this correlation is significant for Negative PEs but not for Positive PEs (average $r > 0$; two-tailed t-tests). H-J. None of the correlations were significant for the second PCA component (PC2). ***$p<0.001$, **$p<0.01$, *p<0.05, ns not significant ($p>0.05$).}
\label{fig:probMatchFig5}
\end{figure}

\paragraph{ROI-based analysis} 
The interpolation parameter $\lambda$ determines to what extent negative feedback is integrated during the learning process, and reinforcement learning theory posits that learning occurs when a prediction error is elicited (i.e. when there is a mismatch between a predicted and an expected feedback; \citep{sutton2018reinforcement}). Together, these results suggest that participants with larger $\lambda$’s (i.e. individuals more inclined to express probability-matching) should show increased neuronal coding of PEs elicited by negative feedback, i.e. negative PEs. Based on previous findings \citep{aberg2015hemispheric}, we tested for group differences in the neuronal correlates of positive and negative prediction errors (PEs) in an a-priori defined VTA ROI by extracting average beta parameter estimates for voxels contained within the VTA ROI (Figure~\ref{fig:probMatchFig5}A).
The beta estimates were entered into a repeated measures ANOVA with within-subject factor PE-type (Positive, Negative) and between-subject factor Group (Maximizing, Probability-matching) based on the outcome of the K-means clustering approach (see above). The ANOVA revealed that the intercept term was significantly different from 0 [$F(1, 55)=5.702, p=0.020$], indicating that the BOLD signal in the VTA ROI correlated significantly with the PE signal across all factors. By contrast, the main effect of PE-type was not significant [$F(1, 55)=0.151, p=0.699$], suggesting no difference between Positive and Negative PEs across the Group factor. Follow-up one-sample t-tests revealed that both types of PEs correlated significantly with BOLD signal across Group [Figure~\ref{fig:probMatchFig5}B; Positive PE: $t(56)=3.445, p=0.001$; Negative PE: $t(56)=2.304, p=0.025$]. The ANOVA revealed a significant main effect of Group [$F(1, 55)=7.601, p=0.008$], but also a significant interaction Group x PE-type [$F(1, 55)=5.763, p=0.020$]. Follow-up two-sample t-tests revealed that the Probability-matching group showed significantly stronger correlation between BOLD signal and Negative PEs as compared to the Maximizing group [Figure~\ref{fig:probMatchFig5}C; $t(55)=2.994, p=0.004$], while the groups did not differ for Positive PEs [Figure~\ref{fig:probMatchFig5}C; $t(55)=0.898, p=0.373$]. These results corroborate the modeling analysis suggesting increased integration of negative outcomes in the learning process of participants in the Probability-matching group (i.e. larger interpolation parameter $\lambda$). No other main effects or interactions were significant (all p-values$>0.05$). \\

\begin{figure}[H]
\centering
\includegraphics[scale=0.8]{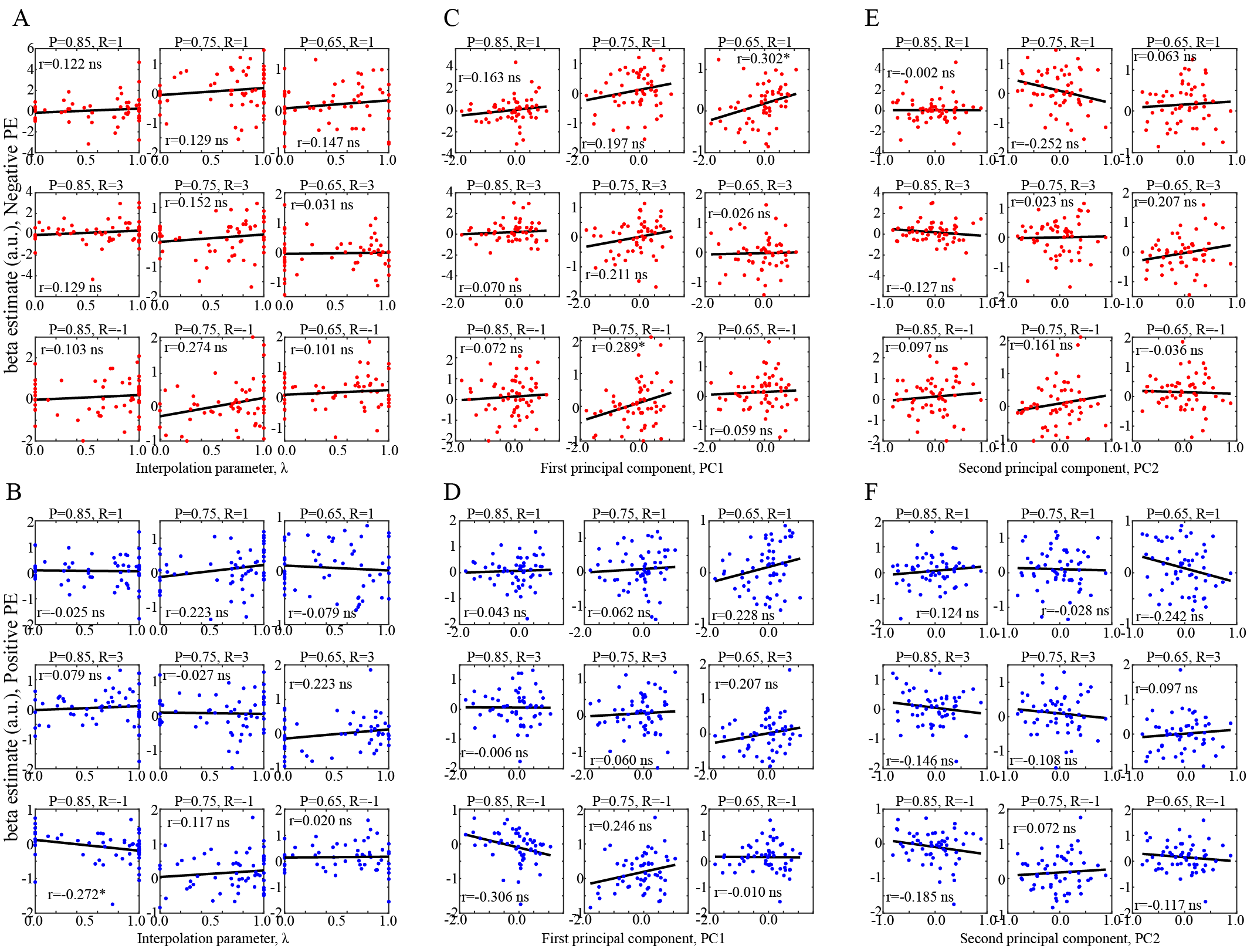}
\caption{Session-wise correlations between the neuronal representation of Negative and Positive PEs in the VTA and the interpolation parameter $\lambda$, or the first and second principal components PC1, PC2. A-B. Correlations between session-specific interpolation parameters $\lambda$ and the VTA representation of Negative (A) and Positive (B) PEs in each session. C-D. Correlations between PC1 and the VTA representation of Negative (C) and Positive (D) PEs in each session. E-F. Correlations between PC2 and the VTA representation of Negative (E) and Positive (F) PEs in each session. *$p<0.05$, ns not significant ($p>0.05$).}
\label{suppFig:pm_supp6}
\end{figure}

In further support of these results, we correlated session-specific interpolation parameters $\lambda$ with the extracted beta estimates for each of the nine conditions, and found that, on average, VTA BOLD signal for Negative PEs, but not Positive PEs, correlated significantly with the $\lambda$’s [Figure~\ref{fig:probMatchFig5}D; average Pearson’s r for Negative PEs$=0.132$, $p<0.001$, $95\%$ CI: $0.083, 0.182$; for Positive PEs$=0.029$, $p=0.595$, $95\%$ CI: $-0.091, 0.148$, two-tailed t-tests; see also Supplementary Figure~\ref{suppFig:pm_supp6}A,B for individual correlations]. \\

In a complementary analysis, the between-subject factor Group in the repeated measures ANOVA was replaced with PC1 as a continuous covariate, i.e. the first PC obtained from the PCA analysis. As would be predicted, all effects remained largely identical [Intercept: $F(1, 55) = 12.428, p<0.001$, PC1: $F(1,55) = 8.240, p=0.005$, PE-type: $F(1,55) = 0.184, p=0.669$, PE-type x PC1: $F(1, 55) = 3.889, p=0.054$]. Follow-up Pearson correlations show that PC1 did not correlate with the representation of Positive PEs [Figure~\ref{fig:probMatchFig5}D; $r=0.181, p=0.177$], but showed a significant and positive correlation with Negative PEs [Figure~\ref{fig:probMatchFig5}E; $r=0.357, p=0.006$]. Session-specific Pearson’s correlation coefficients between PC1 and VTA BOLD signal for Negative and Positive PEs are shown in Figure~\ref{fig:probMatchFig5}G [average Pearson’s r for Negative PEs=$0.154$, $p=0.002$, $95\%$ CI: $0.075, 0.233$; for Positive PEs=$0.058$, $p=0.331$, $95\%$ CI: $-0.072, 0.188$, two-tailed t-tests; see also Supplementary Figure~\ref{suppFig:pm_supp6}C,D for individual correlations]. By contrast, the second PC (PC2) did not correlate significantly with either PE type (Figure~\ref{fig:probMatchFig5}H-J; both p-values $> 0.22$; two-tailed t-tests; see also Supplementary Figure~\ref{suppFig:pm_supp6}E,F for individual correlations). \\

In summary, these results provide evidence that the difference between maximization and probability-matching strategies is related to the midbrain representation of negative prediction errors. \\

\paragraph{Whole-brain analysis} 
Prediction errors correlate with neuronal activity also in brain regions other than the VTA \citep{garrison2013prediction} and negative and positive PEs might be differently represented across the brain \citep{garrison2013prediction, meder2016chasing, pessiglione2006dopamine}. For these reasons, we looked for significant correlations with PEs across the whole-brain using an initial search threshold of $p=0.0001$ and controlling for multiple comparisons using a family-wise error rate for the whole-brain. \\

\begin{figure}[H]
\centering
\includegraphics[scale=0.8]{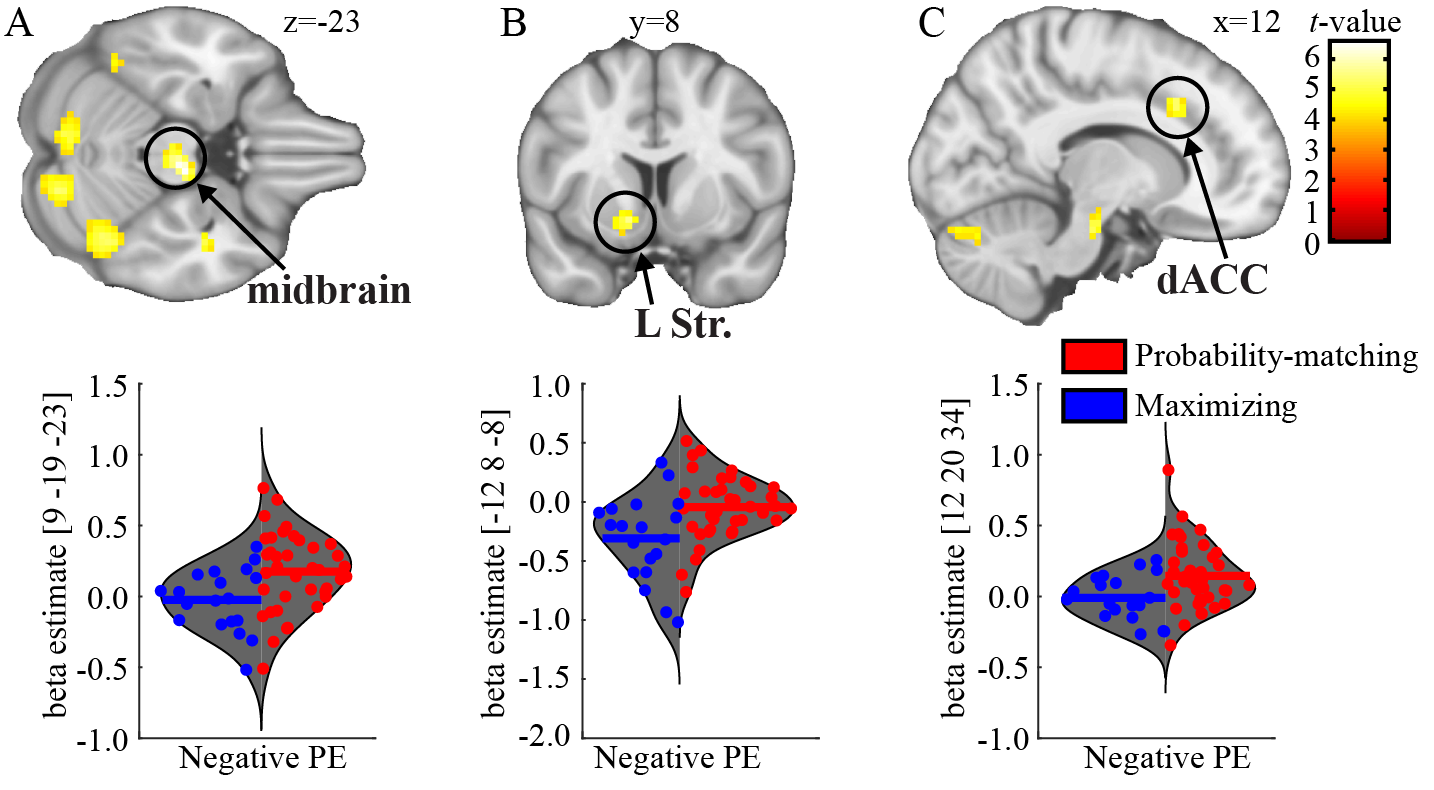}
\caption{Group differences in fMRI correlates of Negative PEs. Participants in the probability-matching group showed stronger correlation between BOLD signal and negative prediction errors in the midbrain (A), the left ventral striatum (B), and in the dorsal anterior cingulate cortex (dACC; C). For illustration purposes, the violin plots display the beta estimate of the peak voxel activation within each cluster of activation. Horizontal lines indicate the mean within each group, while each participant is indicated by a dot. The brain maps are thresholded at $p=0.0001$, and each highlighted cluster was significant when controlling the family-wise error rate for the whole-brain (pFWER$<0.05$).}
\label{fig:probMatchFig6}
\end{figure}

Complementing the conservative VTA ROI analysis of the previous section, participants in the Probability-matching (vs. Maximizing) group showed a stronger correlation between Negative PEs and BOLD signal in a midbrain region (which overlaps with the previous VTA ROI; Figure~\ref{fig:probMatchFig6}A). Extending these results, we also observed stronger correlations between BOLD signal and Negative PEs for the Probability-matching (vs. Maximizing) group in brain regions known to interact with the VTA, such as the left ventral striatum (Figure~\ref{fig:probMatchFig6}B) and the dorsal anterior cingulate cortex (Figure~\ref{fig:probMatchFig6}C). For a full list of brain regions showing stronger correlations between BOLD signal and Negative PEs for the Probability-matching (vs. Maximizing) groups, see Table~\ref{table:t3}. By contrast, no brain region showed stronger encoding of Negative PEs for the Maximizing (vs. Probability-matching) group, nor was there any group differences for Positive PEs (all FWER-corrected p-values$>0.05$).

In summary, the present results confirm predictions derived from the computational model, namely that participants more inclined towards a probability-matching learning strategy show an increased integration of negative outcomes during learning, as indicated by a weaker correlation between BOLD signal and Negative PE signals in brain regions that code PEs.

\begin{figure}[H]
\centering
\includegraphics[scale=0.8]{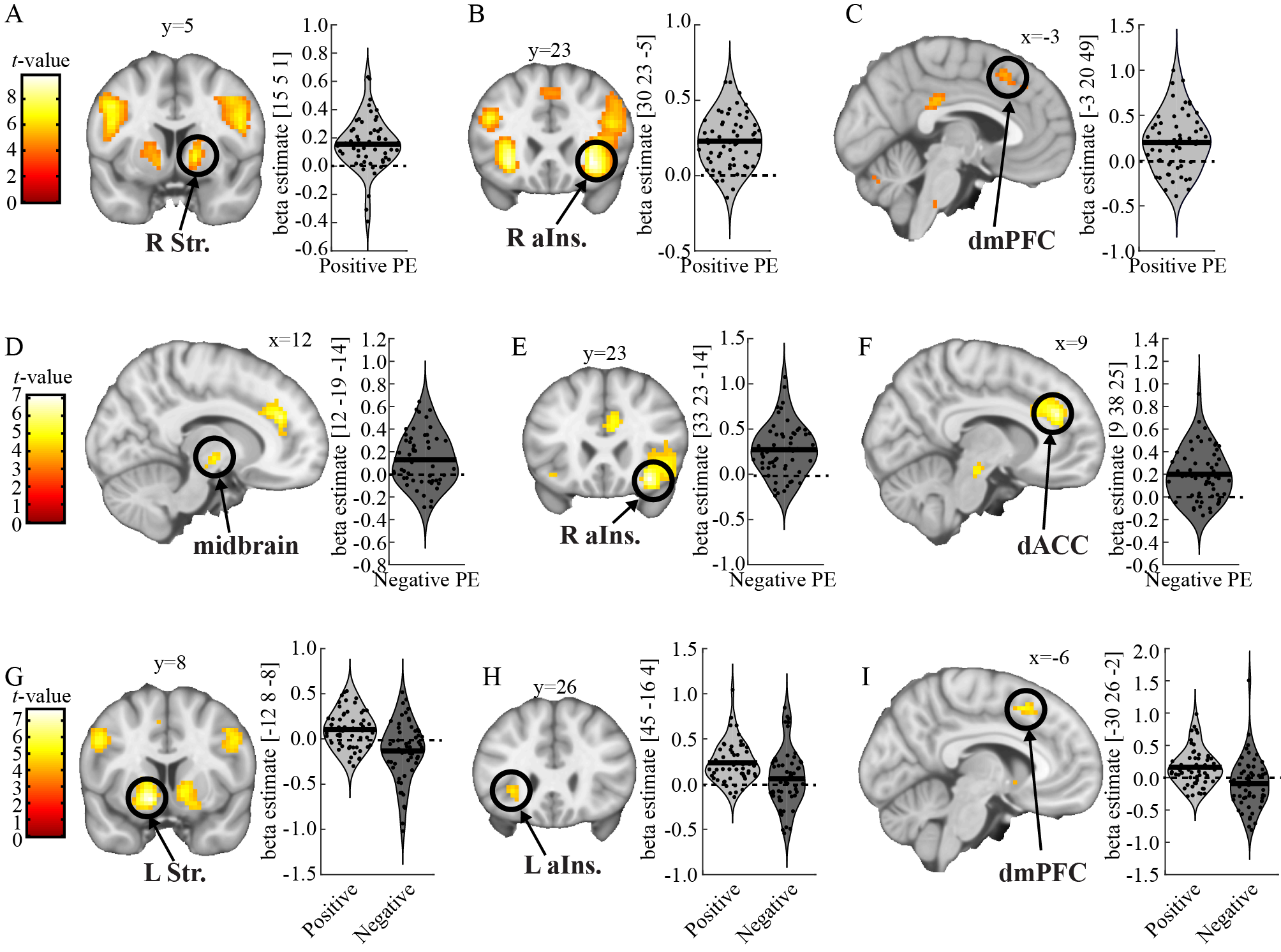}
\caption{Correlations between BOLD signal and PEs. Positive PEs correlated significantly with BOLD signal in the striatum (A), the anterior insula (B), and the dorsomedial prefrontal cortex (C). Negative PEs correlated significantly with BOLD signal in the midbrain (D), the anterior insula (E), and the dorsal anterior cingulate cortex (F). Positive (vs. Negative) PEs showed a stronger correlation with BOLD signal in the striatum (G), the anterior insula (H), and the dorsomedial prefrontal cortex (I). For illustration purposes, the violin plots display the beta estimate of the peak voxel activation within each cluster of activation. Horizontal lines indicate the mean within each PE-type, while each participant is indicated by a dot. The brain maps are thresholded at a $p=0.0001$, and each highlighted cluster was significant when controlling the family-wise error rate for the whole-brain. R = Right, L = Left, Str. = Striatum, aIns. = anterior insula, dmPFC = dorsomedial prefrontal cortex, dACC = dorsal anterior cingulate cortex.}
\label{suppFig:pm_supp7}
\end{figure}

\begin{table}[H]
\begin{center}
\begin{adjustbox}{width=\textwidth}
\begin{tabular}{||c c c c c c c||} 
 \hline
$\mbox{ }$ & \thead{Cluster size,\\ k} & \thead{$p_{FWER}$,\\ voxel-level} & \thead{MNI coordinate,\\ peak voxel: x} & \thead{MNI coordinate,\\ peak voxel: y} & \thead{MNI coordinate,\\ peak voxel: z} & T-value,
peak voxel \\ [0.5ex] 
\hline\hline
\emph{\textbf{Negative PE:}} & \mbox{ } & \mbox{ } & \mbox{ } & \mbox{ } & \mbox{ } & \mbox{ } \\
\hline
\makecell{\textbf{Probability matching} \\ \textbf{$>$ Maximizing}} & \mbox{ } & \mbox{ } & \mbox{ } & \mbox{ } & \mbox{ } & \mbox{ } \\
\hline 
Midbrain & 105 & $< 0.001$ & 9 & -19 & -23 & 6.548 \\
\hline
L Thalamus & 120 & 0.007 & -15 & -16 & 10 & 5.698 \\
\mbox{ } & \mbox{} & 0.035 & -12 & -4 &	4 & 5.169 \\
\hline
L Cerebellum/Fusiform Gyrus & 105 &	0.008 & -45 & -52 & -32 & 5.648 \\
\hline
R Cerebellum/Fusiform Gyrus & 283 &	0.009 & 42 & -55 & -26 & 5.606 \\
\mbox{ } & \mbox{ } & 0.013 & 39 & -64 & -41 & 5.489 \\
\mbox{ } & \mbox{ } & 0.022 & 18 & -79 & -23 & 5.322 \\
\hline
L Cerebellum & 58 &	0.013 & -9 & -73 & -20 & 5.508 \\
\hline
L Ventral Striatum & 64 & 0.013 & -12 &	8 & -8 & 5.491 \\
\hline
R Temporal Lobe & 25 &	0.020 &	42 & -10 & -26 & 5.351 \\
\hline
Dorsal Anterior Cingulate Cortex & 22 &	0.027 &	12 & 20 & 34 & 5.260 \\
\hline
R Middle Temporal Gyrus & 18 &	0.046 &	57 & -58 &	7 & 5.081 \\
\hline
L Middle Frontal Gyrus & 56 & 0.046 & -30 &	50 & 16 & 5.077 \\
\hline
\makecell{\textbf{Maximizing} \\ \textbf{$>$ Probability matching}} & \mbox{ } & \mbox{ } & \mbox{ } & \mbox{ } & \mbox{ } & \mbox{ } \\
\hline 
No significant voxels & \mbox{ } & \mbox{ } & \mbox{ } & \mbox{ } & \mbox{ } & \mbox{ } \\
\hline
\emph{\textbf{Positive PE:}} & \mbox{ } & \mbox{ } & \mbox{ } & \mbox{ } & \mbox{ } & \mbox{ } \\
\hline
\makecell{\textbf{Probability matching} \\ \textbf{$>$ Maximizing}} & \mbox{ } & \mbox{ } & \mbox{ } & \mbox{ } & \mbox{ } & \mbox{ } \\
\hline
No significant voxels & \mbox{ } & \mbox{ } & \mbox{ } & \mbox{ } & \mbox{ } & \mbox{ } \\
\hline
\makecell{\textbf{Maximizing} \\ \textbf{$>$ Probability matching}} & \mbox{ } & \mbox{ } & \mbox{ } & \mbox{ } & \mbox{ } & \mbox{ } \\
\hline 
No significant voxels & \mbox{ } & \mbox{ } & \mbox{ } & \mbox{ } & \mbox{ } & \mbox{ } \\
\hline
\end{tabular}
\end{adjustbox}
  \caption{Group differences in the correlation between BOLD signal and PEs. The initial search threshold was set to $p=0.0001$ and pFWER indicates the p-value after controlling for family-wise error rate (FWER) on the voxel-level across the whole brain.}
\label{table:t3}
\end{center}
\end{table}
Finally, we tested for correlations between Negative and Positive PEs collapsed across all participants. This analysis revealed significant correlations between BOLD signal and Positive PEs in the striatum, anterior insula, and the dorsomedial prefrontal cortex (Supplementary Figure~\ref{suppFig:pm_supp7}A-C). For Negative PEs, we observed significant correlations in the midbrain, the anterior insula, and in the dACC (Supplementary Figure ~\ref{suppFig:pm_supp7}D-F). Moreover, Positive (vs. Negative) PEs showed stronger correlations with BOLD signal in the striatum, the anterior insula, and the dmPFC (Supplementary Figure ~\ref{suppFig:pm_supp7}G-I), while Negative (vs. Positive) PEs showed stronger correlations with BOLD signal in the posterior insula and in the superior frontal gyrus. For a full list and statistics, see Table~\ref{table:t4}.

\begin{table}[H]
\begin{center}
\begin{adjustbox}{width=\textwidth}
\begin{tabular}{||c c c c c c c||} 
 \hline
$\mbox{ }$ & \thead{Cluster size,\\ k} & \thead{$p_{FWER}$,\\ voxel-level} & \thead{MNI coordinate,\\ peak voxel: x} & \thead{MNI coordinate,\\ peak voxel: y} & \thead{MNI coordinate,\\ peak voxel: z} & T-value,
peak voxel \\ [0.5ex] 
\hline\hline
\emph{\textbf{Positive PE:}} & \mbox{ } & \mbox{ } & \mbox{ } & \mbox{ } & \mbox{ } & \mbox{ } \\
\hline 
\makecell{R Anterior Insula \\ / Inferior Frontal Gyrus} & 1267 &	$< 0.001$ & 30 &	23	& -5 & 10.796 \\
\hline
R Middle Frontal Gyrus & \mbox{ } &	$< 0.001$ &	48 & 32 & 31 & 8.296 \\
\hline
R Inferior Frontal Gyrus & \mbox{ } & $< 0.001$ & 48 & 8 & 28 &	7.258 \\
\hline
L Cerebellum & 861 & $< 0.001$ & -36 & -55 & -32 & 9.509 \\
\hline
L Cuneus & \mbox{ } & $< 0.001$ & -12 &	-76	& -26 &	8.877 \\
\hline
L Cerebellum & \mbox{ } & $< 0.001$ & -33 &	-64 & -29 &	8.486 \\
\hline
R Cerebellum & 941 & $< 0.001$ & 33 & -61 &	-29 & 9.418 \\
\hline
R Cerebellum & \mbox{ } & $< 0.001$	& 24 & -70 & -50 & 6.848 \\
\hline
\makecell{L Anterior Insula \\ / Inferior Frontal Gyrus} & 1326 & $< 0.001$	& -30 & 23	& -5 & 9.069 \\
\hline
L Middle Frontal Gyrus & \mbox{ } &	$< 0.001$ &	-45 & 26 & 28 &	7.922 \\
\hline
L Inferior Frontal Gyrus & \mbox{ } & $< 0.001$ & -45 &	5 &	37 & 7.586 \\
\hline
R Striatum / Putamen & 196 & $< 0.001$ & 15 & 5 & 1 & 7.260 \\
\hline
R Thalamus & \mbox{ } &	$0.008$ & 12 & -10 & 7 & 5.670 \\
\hline
Posterior Cingulate Cortex & 91 & $< 0.001$	& 0	& -28 &	31 & 6.399 \\
\hline
L Parietal lobe & 13 & $0.003$ & -36 & -37 & 34 & 5.992 \\
\hline
Supplementary Motor Area & 134 & $0.014$ & -3 &	20 & 49 & 5.492 \\
\hline
Superior Medial Frontal Gyrus & \mbox{ } & $0.028$ & 6 & 29 & 43 & 5.257 \\
\hline
\emph{\textbf{Negative PE:}} & \mbox{ } & \mbox{ } & \mbox{ } & \mbox{ } & \mbox{ } & \mbox{ } \\
\hline
R Superior Frontal Gyrus & 575 & $< 0.001$ & 21 & 50 & 28 &	7.366 \\
\hline
R Dorsal Anterior Cingulate Cortex & \mbox{ } & $< 0.001$ &	9 &	38 & 25 & 6.556 \\
\hline
R Superior Frontal Gyrus & \mbox{ } & $0.010$ & 30 & 44 & 19 & 5.584 \\
\hline
R Inferior Frontal Gyrus & 290 & $< 0.001$ & 33 & 23 & -14 & 7.223 \\
\hline
R Anterior Insula & \mbox{ } & $0.001$ & 36 & 17 & -8 &	6.185 \\
\hline
R Inferior Frontal Gyrus & \mbox{ } & $0.049$ & 48 & 23 & -11 & 5.057 \\
\hline
L Cerebellum & 505 & $0.002$ & -39 & -61 & -29 & 6.081 \\
\hline
L Cerebellum & \mbox{ } & $0.008$ & -39 & -52 & -44 & 5.667 \\
\hline
R Middle Occipital Gyrus & 44 &	$0.019$ & 27 & -94 & -2 & 5.371 \\
\hline
Midbrain & 45 &	$0.027$ & 15 & -10 & -5 & 5.253 \\
\hline
Midbrain & \mbox{ } & $0.060$ & 12 & -19 & -14 & 4.982 \\
\hline
L Anterior Insula & 48 & $0.048$ & -33 & 17 & -2 & 5.063 \\
\hline
\emph{\textbf{Positive $>$ Negative PE:}} & \mbox{ } & \mbox{ } & \mbox{ } & \mbox{ } & \mbox{ } & \mbox{ } \\
\hline
L Putamen / Striatum & 223 & $< 0.001$ & -12 & 8 & -8 &	7.652 \\
\hline
L Insula &  \mbox{ } & $0.079$ & -30 &	26 & -2 & 4.890 \\
\hline
R Putamen / Striatum & 123 & $0.002$ & 15 &	11 & -2 & 6.029 \\
\hline
L Inferior Frontal Gyrus & 132 & $0.004$ & -45 & 5 & 37 & 5.839 \\
\hline
L Supplementary Motor Area & 27 & $0.030$ & -6 & 14 & 49 & 5.226 \\
\hline
R Inferior Frontal Gyrus & 79 & $0.030$ & 48 & 5 & 40 & 5.226 \\
\hline
R Cerebellum & 53 &	$0.034$ & 33 & -64 & -32 & 5.182 \\
\hline
L Middle Frontal Gyrus & 23 & $0.040$ & -39 & 53 & 13 &	5.123 \\
\hline
\emph{\textbf{Negative $>$ Positive PE:}} & \mbox{ } & \mbox{ } & \mbox{ } & \mbox{ } & \mbox{ } & \mbox{ } \\
\hline
R Superior Frontal Gyrus & 193 & $0.004$ & 12 & 53 & 28 & 5.906 \\
\hline
R Superior Frontal Gyrus & \mbox{ } & $0.004$ & 21 & 53 & 31 & 5.872 \\
\hline
R Posterior Insula & 34 & $0.016$ &	45 & -16 & 4 & 5.433 \\
\hline
\end{tabular}
\end{adjustbox}
  \caption{Correlation between BOLD signal and PEs across groups. Initial search threshold was set to $p=0.0001$. pFWER indicates corrected p-value after controlling for family-wise error rate (FWER) on the voxel-level across the whole brain.}
\label{table:t4}
\end{center}
\end{table}

\section{Discussion}
We present a model capable of explaining the full range of learning strategies from pure maximization (i.e. always selecting the ‘best’ option) to full probability matching (i.e. selecting an option according to its probability of providing the ‘best’ outcome). The model was validated via simulations and theoretical proofs, after which it was fitted to actual human performance data in a probabilistic reinforcement learning task. The model-parameter $\lambda$, which determines the balance between the maximization and probability-matching learning strategies, was found to be consistent within an individual across nine different task conditions. This result suggests that the expression of a particular learning strategy is a trait-like feature of an individual. Furthermore, model-based fMRI confirmed predictions derived from the model, namely that individuals more inclined to express probability-matching showed increased integration of negative outcomes, as indicated by a stronger coupling between BOLD signal and negative prediction errors.
The expression of a particular learning strategy, i.e. the degree of maximization (versus probability-matching), showed a trait-like consistency across different task conditions, and was related to the integration of negative outcomes during learning. While personality traits were not assessed here, previous studies report that the relative ability to learn from negative (versus positive) outcomes was related to an individuals' expression of traits estimating sensitivity to punishment and reward \citep{carl2016linking, aberg2017trial}, and high trait anxiety has consistently been associated with increased learning from negative and aversive outcomes \citep{aylward2019altered, huang2017computational, wise2020associations}. Furthermore, it has been suggested that the development and maintenance of some psychiatric symptoms relate to abnormal learning processes \citep{lenaert2014aversive, wise2020associations}, and certain personality traits may indicate an increased vulnerability thereof. It is therefore tempting to suggest that abnormal learning strategies, as (partly) determined by an individual’s personality traits, underpin a biased acquisition of positive and negative information that contributes to dysfunctional behaviors and beliefs. For example, an anxious individual may express a learning strategy that promotes learning about potential dangers in the world and which therefore gradually renders the world increasingly dangerous and hostile, eventually leading to dysfunctional avoidance behaviors. 
As such, interventions aimed at normalizing learning strategies could have beneficial effects on clinical psychiatric symptoms. Notably, it was recently suggested that treatments aimed at altering prediction error signals, for example via pharmacological manipulations of dopamine function, may have beneficial effects on the outcomes of psychotherapy, which itself can be regarded as a form of error-based learning in which dysfunctional beliefs and behaviors are confronted and corrected \citep{papalini2020dopamine}. Supporting this idea, increases in dopaminergic tone was found to be beneficial for learning from positive outcomes (at the expense of learning from negative outcomes), while the reverse effect was found for decreases in dopaminergic tone \citep{bodi2009reward, frank2004carrot}. Yet, further study is clearly needed to determine whether an individual’s learning strategies can be altered, and whether such manipulations may affect psycho-therapeutic outcomes. 
The result that more maximizing participants displayed weaker neuronal coding of negative PEs in the midbrain (as well as in other brain regions) corroborates the model-derived prediction that a maximization learning strategy depends on a reduced integration of negative outcomes during learning. Moreover, this result also lends support to an increasing number of studies reporting that inter-individual differences in learning relate to the neuronal encoding of PEs. For example, \citep{schonberg2007reinforcement} reported that participants that improved performance in a difficult reinforcement learning task showed stronger correlations between PEs and BOLD signal in the striatum, as compared to non-learners. Extending this result, we previously reported that relatively better learning of specific types of information related to relatively stronger correlations between their respective PEs and BOLD signal in the striatum and the midbrain \citep{aberg2015hemispheric, aberg2016left}. Naively, these results make sense because a more efficient neuronal encoding of PEs should lead to better learning, yet findings linking inter-individual differences in learning to the neuronal coding of PEs are surprisingly scarce. Unveiling the neuronal correlates of inter-individual learning abilities and biases is important for understanding the mechanisms of learning in general, and may inform the link between biased learning strategies and psychopathology. 
A pure maximization learning strategy is optimal in reinforcement learning tasks with static outcome probabilities, but is clearly sub-optimal in more dynamic settings where reward contingencies change across trials. Some evidence suggests that learning strategies change as a function of the volatility of the environment, e.g. the optimal learning rate in highly volatile settings is large while smaller learning rates are optimal in more stable environments \citep{behrens2007learning}. The present study offers an alternative explanation to this account, namely that participants may switch between a maximizing learning strategy (when volatility levels are low) and a probability-matching strategy (when volatility is large). Some support for this notion can be obtained from the finding that maximization was more likely when the uncertainty was low (i.e. when the outcome Probability was high; P=0.85). Future studies should address to what extent learning-strategies can be flexibly adjusted in a dynamic environment, and whether such models provide a better description of behavior as compared to models with adaptive learning rates. 
Our results provide insights into the concept of probability-matching, which is often explained via higher order mechanisms \citep{vulkan2000economist}. For example, previous research suggests that people look for patterns in a feedback sequence even if they are told that the rewards are sampled randomly from the distribution, and that people do not have an understanding of the concept of randomness \citep{gaissmaier2008smart}. This idea is supported by studies reporting that when people could see the random device selecting the outcome they tended to employ more of a maximizing strategy, as compared to people that could not see the device \citep{morse1960probability}. Another suggestion is that predicting infrequent events gives higher utility than predicting frequent events, thus people do not converge to a maximization strategy \citep{brackbill1962supplementary}. As mentioned in the Introduction, probability-matching has been found in non-human species, something that questions to what extent higher order cognitive processes are necessary to express probability-matching. Our study provides evidence that probability-matching is a consequence of a lower-level mechanism, namely an increased integration of negative outcomes during learning. Importantly, this conclusion is supported by both theoretical proofs and neuroimaging results which are well-grounded in reinforcement learning theory. Importantly, our model should be seen as complementary to previous research rather than replacing it. Particularly interesting for future research is to determine whether learning strategies are under top-down control, including the neuronal encoding of PEs, a notion that would allow merging our findings with previous research. \\

Notably, while probability-matching is suboptimal during learning with static reward probabilities, it reduces the uncertainty of the ‘worst’ options (via an increased sampling rate) and may therefore be beneficial for learning about the decision making environment in general. Supporting this idea, we recently showed that participants showing relatively stronger midbrain coding of negative (versus positive) PEs during a learning phase, reminiscent of a probability-matching learning strategy in the present study, showed better discrimination performance between the ‘worst’ options in a post-learning test phase \citep{carl2016linking}. To clarify, in the initial learning phase, participants learned to select the option with the highest reward probability in three separate pairs of objects (reward probabilities for the ‘best’/’worst’ object in a pair were respectively: 0.8/0.2, 0.7/0.3, and 0.6/0.4). In the subsequent test phase, conducted without feedback to prevent further learning, these objects were mixed to create new pairs (e.g. the ‘best’ pairs: 0.8/0.7, 0.8/0.6, 0.7/0.6; the ‘worst’ pairs: 0.2/0.3, 0.2/0.4, 0.3/0.4). Participants showing relatively stronger midbrain coding of negative (versus positive) PEs during learning were subsequently relatively better at discriminating between the ‘worst’ (versus the ‘best’) pairs. One interpretation of these results is that the differential midbrain coding of positive and negative PEs influenced to what extent positive and negative outcomes were integrated during the learning. Yet, given the present results, an alternative explanation is that probability-matching increased the sampling of the ‘worst’ options, leading to a less uncertain representation of their expected values, and therefore also better subsequent value-based discrimination performance. \\

A related and noteworthy contribution of the present thesis is that the exploitation/exploration trade-off can be understood as a learning bias rather than a decision bias. For example, most human neuroscience literature regards the exploitation/exploration trade-off as a decision strategy, where exploitation is aimed at maximizing immediate value gains, while exploration is promoted by the possibility to reduce uncertainty in the decision making environment \citep{badre2012rostrolateral, chakroun2020dopaminergic, tomov2020dissociable, wilson2021balancing}. In other words, when facing exactly the same choice, an ‘exploiter’ assigns a higher utility to possible value gains, while an ‘explorer’ emphasize gains in information. By contrast, our results suggest that an ‘exploiter’ could also be an individual that expresses a maximization learning strategy, while an ‘explorer’ engages more in probability-matching. Accordingly, the difference between an ‘exploiter’ and an ‘explorer’ may not only be determined by the subjective utilities of the different factors available for a decision (e.g. expected values, risk, potential information gains), but these factors may be biased due to preceding learning strategies that emphasize learning of a particular type of information. Put simply, an individual unable to learn from negative outcomes (e.g. bad dishes served in a restaurant or a quarreling partner), should be more likely to remain and exploit the positive aspects of those options (e.g. good deserts or a loving partner), while an ‘explorer/probability-matcher’ should be more likely to try out other alternatives. 

\newpage
\bibliography{biblio_final_report}
\bibliographystyle{plainnat}

\newpage
\section{Appendix}
\label{sec:appendixA}
\subsection{Theoretical Analysis}
We will show in this section for each one of the models, that it converges in the limit of infinite trials, and that the convergence is to a fixed distribution of the choice probabilities. This means that we do not have a distribution that changes indefinitely, rather it converges to a stable choice distribution, which is the chosen "strategy". We start with a warm-up in section \ref{fixedPoints} to give intuition for the formal proof in section \ref{convergenceAnalysis}. In this section we compute what are the fixed points of the models, if in every iteration probabilities would have been updated by the expected value. This will give intuition for why the model converges to the fixed points that we find, as proved formally in section \ref{convergenceAnalysis}.

\subsubsection{Fixed Points Analysis}
\label{fixedPoints}
\paragraph{Probability-matching model} \mbox{ }  \\
We will give an analysis for the fixed points of the learning rule both in the full information case, or in the full and partial information case when $K=2$.
In these settings we have the model:
\begin{gather*}
\mbox{For } i=1,2...,K \mbox{ and } 0 \leq \alpha \leq 1: \\
p_{i,n+1}=\begin{cases}
p_{i,n}+\alpha\cdot(1-p_{i,n}) & if\,\, X_{i,n}=1\\
p_{i,n}-\alpha\cdot p_{i,n} & otherwise
\end{cases}
\end{gather*}
We will compute the expected value of $p_{i,n+1}$ given $p_{i,n}$:\\
\begin{gather*}
\mathbb{E}[p_{i,n+1}|p_{i,n}]=p_{i,n}+\alpha\cdot(1-p_{i,n})\cdot q_{i}-\alpha\cdot p_{i,n}\cdot(1-q_{i})= \\
=p_{i,n}+\alpha\cdot(q_{i}-p_{i,n}\cdot q_{i}-p_{i,n}+p_{i,n}\cdot q_{i})=p_{i,n}+\alpha\cdot(q_{i}-p_{i,n})
\end{gather*}
A fixed point is a point for which: $p_{i,n+1}=p_{i,n}$, this happens here when:\\
$q_{i}-p_{i,n}=0\iff q_{i}=p_{i,n}$\\
This is a stable fixed point:\\
If $p_{i,n}>q_{i}$: $q_{i}-p_{i,n}<0\,\Longrightarrow p_{i,n+1}<p_{i,n}$ \\
If $p_{i,n}<q_{i}$: $q_{i}-p_{i,n}>0\,\Longrightarrow p_{i,n+1}>p_{i,n}$\\
$\implies $ Fixed points: $p_{i,n}=q_{i} \implies $ converges to probability matching.\\

\paragraph{Maximization model}  \mbox{ } \\
We first remind the reader the model:\\
\begin{gather*}
\mbox{For } i \in \{1,2,...,K\} \mbox{ and } 0 \leq \alpha \leq 1:\\
p_{i,n+1}=\begin{cases}
p_{i,n}+\alpha\cdot(1-p_{i,n}) & if\,\, j(n)=i\,\, and\,\, X_{i,n}=1\\
p_{i,n} & if\,\, j(n)=i\,\, and\,\, X_{i,n}=0\\
p_{i,n}-\alpha\cdot p_{i,n} & if\,\, j(n)\neq i\,\, and\,\, X_{j(n),n}=1\\
p_{i,n} & if\,\, j(n)\neq i\,\, and\,\, X_{j(n),n}=0
\end{cases}
\end{gather*}
We will compute the expected value of $p_{i,n+1}$ given $p_{i,n}$:
\begin{gather*}
\mathbb{E}[p_{i,n+1}|p_{i,n}]=p_{i,n}+\alpha\cdot(1-p_{i,n})\cdot p_{i,n}\cdot q_{i}-\alpha\cdot p_{i,n}\cdot\underset{j\neq i}{\sum}p_{j,n}\cdot q_{j}= \\
=p_{i,n}+\alpha\cdot(p_{i,n}\cdot q_{i}-p_{i,n}\cdot(p_{i,n}\cdot q_{i}+\underset{j\neq i}{\sum}p_{j,n}\cdot q_{j}))= \\
=p_{i,n}+\alpha\cdot(p_{i,n}\cdot q_{i}-p_{i,n}\cdot\underset{j=1}{\overset{k}{\sum}}p_{j,n}\cdot q_{j})= \\
=p_{i,n}+\alpha\cdot(p_{i,n}\cdot(q_{i}-\underset{j=1}{\overset{k}{\sum}}p_{j,n}\cdot q_{j}))= \\
=p_{i,n}\cdot(1+\alpha\cdot(q_{i}-\underset{j=1}{\overset{k}{\sum}}p_{j,n}\cdot q_{j}))= \\
=p_{i,n}\cdot(1+\alpha\cdot(q_{i}\cdot(1-p_{i,n})-\underset{j\neq i}{\sum}p_{j,n}\cdot q_{j}))= \\
=p_{i,n}\cdot(1+\alpha\cdot(q_{i}\cdot\underset{j\neq i}{\sum}p_{j,n}-\underset{j\neq i}{\sum}p_{j,n}\cdot q_{j}))= \\
=p_{i,n}\cdot(1+\alpha\cdot(\underset{j\neq i}{\sum}p_{j,n}\cdot(q_{i}-q_{j})))
\end{gather*}
A fixed point is a point for which: $p_{i,n+1}=p_{i,n}$, this happens here when: \\
$ p_{i,n}=0\,\Longrightarrow\mathbb{E}[p_{i,n+1}|p_{i,n}]=0\cdot(1+\alpha\cdot(q_{i}-\underset{j=1}{\overset{k}{\sum}}p_{j,n}\cdot q_{j}))=0 $ \\
or: \\
$ p_{i,n}=1\,\Longrightarrow\mathbb{E}[p_{i,n+1}|p_{i,n}]=1\cdot(1+\alpha\cdot0)=1 $ \\
or: \\
$ p_{i,n}=1-\frac{\underset{j\neq i}{\sum}p_{j,n}\cdot q_{j}}{q_{i}}\,\Longrightarrow q_{i}-\underset{j=1}{\overset{k}{\sum}}p_{j,n}\cdot q_{j}=0 $ \\
Note that for the machine with the highest winning probability, we get that: \\
$ \alpha\cdot(\underset{j\neq i}{\sum}p_{j,n}\cdot\underset{\geq0}{\underbrace{(q_{i}-q_{j})}}))\geq0\,\implies\, p_{i,n+1}\geq p_{i,n} $. \\
So we conclude that this model converges to maximization.

\subsubsection{Convergence Analysis}
\label{convergenceAnalysis}
\paragraph{Probability-matching model} \mbox{ } \\
Let $ p_{i,n+1} $ be the probability of the learner to choose machine $i$ in trial $n+1$, i.e. $p_{i,n+1}=pr(A_{i,n+1})$. \\
Then, for $i=1,2$ and $0 \leq \alpha \leq 1$:\\
\begin{gather*}  
p_{i,n+1}=\begin{cases}
p_{i,n}+\alpha\cdot(1-p_{i,n}) & if\,\, X_{i,n}=1\\
p_{i,n}-\alpha\cdot p_{i,n} & otherwise
\end{cases}
\end{gather*}
We will calculate the expectation of the probability of choosing machine $i$, under these update rules:\\
We compute the expectation of $p_{i,n}$: $\mathbb{E}[p_{i,n}]$. We note that at every round we get further information on the variable $p_{i,n}$ and thus from the law of iterated expectation we have: \\
\begin{gather*}
\mathbb{E}[p_{i,n}]=\mathbb{E}[\mathbb{E}[p_{i,n}|p_{i,1}]]=\mathbb{E}[\mathbb{E}[\mathbb{E}[p_{i,n}|p_{i,1},p_{i,2}]]]=...=\underset{n-1\,\, expectations}{\underbrace{\mathbb{E}[\mathbb{E}[...\mathbb{E}[p_{i,n}|p_{i,1},p_{i,2},...,p_{i,n-1}]]]}}
\end{gather*} 
Note that: \\
\begin{gather*}
\mathbb{E}[p_{i,n}|p_{i,n-1},p_{i,n-2},...,p_{i,1}]= \\
=(p_{i,n-1}+\alpha\cdot(1-p_{i,n-1}))\cdot\underset{denote:\, q_{i}}{\underbrace{Pr(X_{i,n}=1)}}+(p_{i,n-1}-\alpha\cdot p_{i,n-1})\cdot\underset{1-q_{i}}{\underbrace{(1-Pr(X_{i,n}=1))}}= \\
=q_{i}\cdot(1-\alpha)\cdot p_{i,n-1}+\alpha\cdot q_{i}+(1-\alpha)\cdot p_{i,n-1}-q_{i}\cdot(1-\alpha)\cdot p_{i,n-1}=\alpha\cdot q_{i}+(1-\alpha)\cdot p_{i,n-1}
\end{gather*}
We continue by iterating over the expectations and get: \\
\begin{gather*}
\mathbb{E}[p_{i,n}]=\alpha\cdot q_{i}\cdot\underset{r=1}{\overset{n}{\sum}}(1-\alpha)^{n-r}+(1-\alpha)^{n-1}\cdot p_{i,0}
\end{gather*}
We note that $(1-\alpha)<1$ and so:\\
\begin{gather*}
\underset{n\rightarrow\infty}{lim}\mathbb{E}^{0}[p_{i,n}]=\underset{n\rightarrow\infty}{lim}\alpha\cdot q_{i}\cdot\underset{r=1}{\overset{n}{\sum}}(1-\alpha)^{n-r}=\alpha\cdot q_{i}\cdot\frac{1}{1-(1-\alpha)}=q_{i}
\end{gather*}
So what we get is that in this full information with a linear reinforcement model, the convergence of the expected probability to choose machine i, is the probability of getting rewarded in machine i, i.e, we converge to probability matching. \\
We will show now that not only the expected probabilities to choose each machine converge to probability matching, but that the probability to choose machine i at every time stamp $t$ converges to probability matching, i.e. we will look at the variance of $p_{i,n}$ as $n$ goes to infinity and show that: \\
\begin{gather*}
\forall\epsilon>0,\,\,\exists\alpha:\,\,0\leq\alpha\leq1\,\, and\,\,\underset{n\rightarrow\infty}{lim}\mathbb{E}[(p_{i,n}-\mathbb{E}[p_{i,n}])^{2}]\leq\epsilon
\end{gather*}
We start by getting a convenient expression for: $\mathbb{E}[(p_{i,n}-\mathbb{E}[p_{i,n}])^{2}]$
$\mathbb{E}[(p_{i,n}-\mathbb{E}[p_{i,n}])^{2}]=\mathbb{E}[p_{i,n}^{2}]-(\mathbb{E}[p_{i,n}])^{2}$ \\
We now check what is: $\mathbb{E}[p_{i,n}^{2}]$: \\
$\mathbb{E}[p_{i,n}^{2}]=\mathbb{E}[\mathbb{E}[...\mathbb{E}[p_{i,n}^{2}|p_{i,1},p_{i,2},...,p_{i,n-1}]]]$ \\
Note that: \\
\begin{gather*}
\mathbb{E}[p_{i,n}^{2}|p_{i,1},p_{i,2},...,p_{i,n-1}]= \\
= (p_{i,n-1}+\alpha\cdot(1-p_{i,n-1}))\cdot q_{i}+(p_{i,n-1}-\alpha\cdot p_{i,n-1})\cdot(1-q_{i})\,)^{2}= \\
= (\alpha\cdot q_{i}+p_{i,n-1}-\alpha\cdot p_{i,n-1})^{2}= \\
=(p_{i,n-1}+(\alpha\cdot q_{i}-\alpha\cdot p_{i,n-1}))^{2}=p_{i,n-1}^{2}+2p_{i,n-1}\cdot(\alpha\cdot q_{i}-\alpha\cdot p_{i,n-1})+(\alpha\cdot q_{i}-\alpha\cdot p_{i,n-1})^{2}= \\
=p_{i,n-1}^{2}+2p_{i,n-1}\cdot\alpha\cdot q_{i}-2\cdot\alpha\cdot p_{i,n-1}^{2}+\alpha^{2}\cdot q_{i}^{2}-2\alpha^{2}\cdot q_{i}\cdot p_{i,n-1}+\alpha^{2}\cdot p_{i,n-1}^{2}= \\
=(2\cdot\alpha\cdot q_{i}-2\alpha^{2}\cdot q_{i})\cdot p_{i,n-1}+(1-2\alpha+\alpha^{2})\cdot p_{i,n-1}^{2}+\alpha^{2}\cdot q_{i}^{2}
\end{gather*}
So we get: \\
\begin{gather*}
\mathbb{E}[p_{i,n}^{2}]=\mathbb{E}[\mathbb{E}[...\mathbb{E}[p_{i,n}^{2}|p_{i,1},p_{i,2},...,p_{i,n-1}]]]= \\
\mathbb{E}[(2\cdot\alpha\cdot q_{i}-2\theta^{2}\cdot q_{i})\cdot p_{i,n-1}+(1-2\alpha+\alpha^{2})\cdot p_{i,n-1}^{2}+\alpha^{2}\cdot q_{i}^{2}]= \\
\mathbb{E}[(2\cdot\alpha\cdot q_{i}-2\alpha^{2}\cdot q_{i})\cdot p_{i,n-1}]+(1-\alpha)^{2}\cdot \mathbb{E}[p_{i,n-1}^{2}]+\alpha^{2}\cdot q_{i}^{2}
\end{gather*}
We now apply the limit: \\
\begin{gather*}
\underset{n\rightarrow\infty}{lim}\mathbb{E}[p_{i,n}^{2}]=\underset{n\rightarrow\infty}{lim}(\, \mathbb{E}[(2\cdot\alpha\cdot q_{i}-2\alpha^{2}\cdot q_{i})\cdot p_{i,n-1}]+(1-\alpha)^{2}\cdot \mathbb{E}[p_{i,n-1}^{2}]+\alpha^{2}\cdot q_{i}^{2}\,)= \\
=(2\cdot\alpha\cdot q_{i}-2\alpha^{2}\cdot q_{i})\cdot q_{i}+(1-\alpha)^{2}\cdot\underset{n\rightarrow\infty}{lim}\mathbb{E}[p_{i,n-1}^{2}]+\alpha^{2}\cdot q_{i}^{2}= \\
=\underset{n\rightarrow\infty}{lim}\underset{r=1}{\overset{n}{\sum}}((1-\alpha)^{2})^{n-r}\cdot(2\cdot\alpha\cdot q_{i}-2\alpha^{2}\cdot q_{i})\cdot q_{i}+\underset{r=1}{\overset{n}{\sum}}((1-\alpha)^{2})^{n-r}\cdot\alpha^{2}\cdot q_{i}^{2}= \\
=\frac{(2\cdot\alpha\cdot q_{i}-2\alpha^{2}\cdot q_{i})\cdot q_{i}}{1-(1-\alpha)^{2}}+\frac{\alpha^{2}\cdot q_{i}^{2}}{1-(1-\alpha)^{2}}
\end{gather*}
Thus we get: \\
\begin{gather*}
\underset{n\rightarrow\infty}{lim}\mathbb{E}[p_{i,n}^{2}]=\underset{n\rightarrow\infty}{lim}\underset{r=1}{\overset{n}{\sum}}((1-\alpha)^{2})^{n-r}\cdot(2\cdot\alpha\cdot q_{i}-2\alpha^{2}\cdot q_{i})\cdot q_{i}+\underset{r=1}{\overset{n}{\sum}}((1-\alpha)^{2})^{n-r}\cdot\alpha^{2}\cdot q_{i}^{2}= \\
\frac{(2\cdot\alpha\cdot q_{i}-2\alpha^{2}\cdot q_{i})\cdot q_{i}}{1-(1-\alpha)^{2}}+\frac{\alpha^{2}\cdot q_{i}^{2}}{1-(1-\alpha)^{2}}=\frac{q_{i}^{2}\cdot(2\alpha-2\alpha^{2})}{2\alpha-\alpha^{2}}+\frac{\alpha^{2}\cdot q_{i}^{2}}{2\alpha-\alpha^{2}}
\end{gather*}
So we get that the limit of the variance of $p_{i,n}$ is: \\
\begin{gather*}
\underset{n\rightarrow\infty}{lim}Var[p_{i,n}]=\underset{n\rightarrow\infty}{lim}\mathbb{E}[(p_{i,n}-\mathbb{E}[p_{i,n}])^{2}]=\underset{n\rightarrow\infty}{lim}\mathbb{E}[p_{i,n}^{2}]-(\mathbb{E}[p_{i,n}])^{2}= \\
\underset{n\rightarrow\infty}{lim}\mathbb{E}[p_{i,n}^{2}]-\underset{n\rightarrow\infty}{lim}(\mathbb{E}[p_{i,n}])^{2}=\frac{q_{i}^{2}\cdot(2\alpha-2\alpha^{2})}{2\alpha-\alpha^{2}}+\frac{\alpha^{2}\cdot q_{i}^{2}}{2\alpha-\alpha^{2}}-q_{i}^{2}= \\
=\frac{\alpha^{2}\cdot q_{i}^{2}}{2\alpha-\alpha^{2}}+(\frac{2\alpha-2\alpha^{2}}{2\alpha-\alpha^{2}}-1)\cdot q_{i}^{2}=\frac{\alpha\cdot q_{i}^{2}}{2-\alpha}+\frac{-\alpha}{2-\alpha}\cdot q_{i}^{2}\underset{since\,0\leq\alpha\leq1}{\underbrace{\leq}}\frac{\alpha\cdot q_{i}^{2}}{2-\alpha}\underset{since\,0\leq\alpha\leq1}{\underbrace{\leq}}\alpha\cdot q_{i}^{2}\underset{since\,0\leq q_{i}\leq1}{\underbrace{\leq}}\alpha
\end{gather*}

\paragraph{Maximiztion Model}
Our learning rule in this model is:
\begin{gather*}
\mbox{For } i \in \{1,2,...,K\} \mbox{ and }  0 \leq \alpha \leq 1:\\
p_{i,n+1}=\begin{cases}
p_{i,n}+\alpha\cdot(1-p_{i,n}) & if\,\, j(n)=i\,\, and\,\, X_{i,n}=1\\
p_{i,n} & if\,\, j(n)=i\,\, and\,\, X_{i,n}=0\\
p_{i,n}-\alpha\cdot p_{i,n} & if\,\, j(n)\neq i\,\, and\,\, X_{j(n),n}=1\\
p_{i,n} & if\,\, j(n)\neq i\,\, and\,\, X_{j(n),n}=0
\end{cases}
\end{gather*}
We will compute the expected value of $p_{i,n+1}$ given $p_{i,n}$, and then use this to compute the expected value of $p_{i,n} $ as $n$ goes to infinity: \\
\begin{gather*}
\mathbb{E}[p_{i,n+1}|p_{i,n}]=p_{i,n}+\alpha\cdot(1-p_{i,n})\cdot p_{i,n}\cdot q_{i}-\alpha\cdot p_{i,n}\cdot\mathbb{E}[\underset{j\neq i}{\sum}p_{j,n}\cdot q_{j}|p_{i,n}]= \\
=p_{i,n}+\alpha\cdot(p_{i,n}\cdot q_{i}-p_{i,n}\cdot(p_{i,n}\cdot q_{i}+\mathbb{E}[\underset{j\neq i}{\sum}p_{j,n}\cdot q_{j}|p_{i,n}]))= \\
=p_{i,n}+\alpha\cdot(p_{i,n}\cdot q_{i}-p_{i,n}\cdot\mathbb{E}[\underset{j=1}{\overset{k}{\sum}}p_{j,n}\cdot q_{j}|p_{i,n}])= \\
=p_{i,n}+\alpha\cdot(p_{i,n}\cdot(q_{i}-\mathbb{E}[\underset{j=1}{\overset{k}{\sum}}p_{j,n}\cdot q_{j}|p_{i,n}]))= \\
=p_{i,n}\cdot(1+\alpha\cdot(q_{i}-\mathbb{E}[\underset{j=1}{\overset{k}{\sum}}p_{j,n}\cdot q_{j}|p_{i,n}]))= \\
=p_{i,n}\cdot(1+\alpha\cdot(q_{i}\cdot(1-p_{i,n})-\mathbb{E}[\underset{j\neq i}{\sum}p_{j,n}\cdot q_{j}|p_{i,n}]))= \\
=p_{i,n}\cdot(1+\alpha\cdot(q_{i}\cdot\mathbb{E}[\underset{j\neq i}{\sum}p_{j,n}-\underset{j\neq i}{\sum}p_{j,n}\cdot q_{j}|p_{i,n}]))= \\
=p_{i,n}\cdot(1+\alpha\cdot(\mathbb{E}[\underset{j\neq i}{\sum}p_{j,n}\cdot(q_{i}-q_{j})|p_{i,n}]))
\end{gather*}
Assume $i=\underset{j\in\{1,...,K\}}{argmax}q_{j}$ and denote by: $\Delta_{min}=q_{i}-\underset{j\neq i}{max}\{q_{j}\}$ \\
Note that for this particular $i$ we get that $\Delta_{min}>0$ and: \\
\begin{gather*}
\mathbb{E}[p_{i,n+1}|p_{i,n}]=p_{i,n}\cdot(1+\alpha\cdot(\mathbb{E}[\underset{j\neq i}{\sum}p_{j,n}\cdot(q_{i}-q_{j})|p_{i,n}]))= \\
= p_{i,n}+\alpha\cdot p_{i,n}\cdot\mathbb{E}[(\underset{j\neq i}{\sum}p_{j,n}\cdot(q_{i}-q_{j}))|p_{i,n}]\geq p_{i,n}+\alpha\cdot p_{i,n}\cdot\mathbb{E}[(\underset{j\neq i}{\sum}p_{j,n}\cdot\Delta_{min})|p_{i,n}]\geq p_{i,n}
\end{gather*}
Thus we have here a sequence of random variables $p_{i,1},p_{i,2},...,p_{i,n},...$ satisfying: \\ 
$\mathbb{E}[p_{i,n+1}|p_{i,n},p_{i,n-1},...,p_{i,0}]\geq p_{i,n}$ and in addition $\mathbb{E}[|p_{i,n}|]=\mathbb{E}[p_{i,n}]<\infty $.
This is a discrete time submartingale. Thus we know by the submartingale convergence theorem that $p_{i,n}\longrightarrow p_{i,\infty}$ and that $\mathbb{E}[p_{i,n}]\longrightarrow\mathbb{E}[p_{i,\infty}]$. \\
Note that: \\
\begin{gather*}
\mathbb{E}[p_{i,n+1}]=\mathbb{E}[\mathbb{E}[p_{i,n+1}|p_{i,n}]]=\mathbb{E}[p_{i,n}+\alpha\cdot p_{i,n}\cdot(\mathbb{E}[\underset{j\neq i}{\sum}p_{j,n}\cdot(q_{i}-q_{j})|p_{i,n}]))]
\end{gather*}
We now turn to compute this limit, i.e. \\
$\mathbb{E}[p_{i,\infty}]=\underset{n\rightarrow\infty}{lim}\,\mathbb{E}[p_{i,n+1}]=\underset{n\rightarrow\infty}{lim}\,\mathbb{E}[p_{i,n}+\alpha\cdot p_{i,n}\cdot(\mathbb{E}[\underset{j\neq i}{\sum}p_{j,n}\cdot(q_{i}-q_{j})|p_{i,n}]))]$ \\
Since: \\
$\forall n:\,\mathbb{E}[p_{i,n}+\alpha\cdot p_{i,n}\cdot(\mathbb{E}[\underset{j\neq i}{\sum}p_{j,n}\cdot(q_{i}-q_{j})|p_{i,n}]))]\geq\mathbb{E}[p_{i,n}+\alpha\cdot p_{i,n}\cdot\mathbb{E}[(\underset{j\neq i}{\sum}p_{j,n}\cdot\Delta_{min})|p_{i,n}]]$, we get that: \\
\begin{gather*}
\underset{n\rightarrow\infty}{lim}\,\mathbb{E}[p_{i,n}+\alpha\cdot p_{i,n}\cdot(\mathbb{E}[\underset{j\neq i}{\sum}p_{j,n}\cdot(q_{i}-q_{j})|p_{i,n}]))]\geq\underset{n\rightarrow\infty}{lim}\,\mathbb{E}[p_{i,n}+\alpha\cdot p_{i,n}\cdot\mathbb{E}[(\underset{j\neq i}{\sum}p_{j,n}\cdot\Delta_{min})|p_{i,n}]))]= \\
= \underset{n\rightarrow\infty}{lim}\,\mathbb{E}[p_{i,n}+\alpha\cdot p_{i,n}\cdot\Delta_{min}\cdot(1-p_{i,n})))]= \\
= \underset{n\rightarrow\infty}{lim}\,\mathbb{E}[p_{i,n}]+\alpha\cdot\Delta_{min}\cdot\mathbb{E}[p_{i,n}\cdot(1-p_{i,n})]=\underset{n\rightarrow\infty}{lim}\,\mathbb{E}[p_{i,\infty}]+\alpha\cdot\Delta_{min}\cdot\mathbb{E}[p_{i,n}\cdot(1-p_{i,n})]
\end{gather*}
and:
\begin{gather*}
\mathbb{E}[p_{i,\infty}]=\underset{n\rightarrow\infty}{lim}\,\mathbb{E}[p_{i,n+1}]\geq\underset{n\rightarrow\infty}{lim}\,\mathbb{E}[p_{i,\infty}]+\alpha\cdot\Delta_{min}\cdot\mathbb{E}[p_{i,n}\cdot(1-p_{i,n})] \\
\implies0\geq\underset{n\rightarrow\infty}{lim}\,\alpha\cdot\Delta_{min}\cdot\mathbb{E}[p_{i,n}\cdot(1-p_{i,n})]\,\implies\,0\geq\underset{n\rightarrow\infty}{lim}\,\mathbb{E}[p_{i,n}\cdot(1-p_{i,n})]=\underset{n\rightarrow\infty}{lim}\,\mathbb{E}[p_{i,n}-p_{i,n}^{2}]
\end{gather*}
Note that since $ \forall n:\, p_{i,n}\in[0,1]\,\implies\,\underset{n\rightarrow\infty}{lim}\,\mathbb{E}[p_{i,n}\cdot(1-p_{i,n})]\geq 0$. Thus we get: \\
$\underset{n\rightarrow\infty}{lim}\,\mathbb{E}[p_{i,n}\cdot(1-p_{i,n})]=0 $ \\
So now we have: \\
$\underset{n\rightarrow\infty}{lim}\,\mathbb{E}[p_{i,n}-p_{i,n}^{2}]=0$ \\
Multiplying by -1 both sides we get: \\
$\underset{n\rightarrow\infty}{lim}\,\mathbb{E}[p_{i,n}^{2}-p_{i,n}]=0$ \\
and now using Jensen's inequality we get: \\
\begin{gather*}
0=\underset{n\rightarrow\infty}{lim}\,\mathbb{E}[p_{i,n}^{2}-p_{i,n}]\geq\underset{n\rightarrow\infty}{lim}\,(\mathbb{E}^{2}[p_{i,n}]-\mathbb{E}[p_{i,n}])=\mathbb{E}^{2}[p_{i,\infty}]-\mathbb{E}[p_{i,\infty}]=\mathbb{E}[p_{i,\infty}]\cdot(1-\mathbb{E}[p_{i,\infty}])
\end{gather*}
This implies that: $\mathbb{E}[p_{i,\infty}]=0$ or: $ (1-\mathbb{E}[p_{i,\infty}])=0 $, i.e. : \\ $\mathbb{E}[p_{i,\infty}]\in\{0,1\}$ \\
Now since for $i=\underset{j\in\{1,...,K\}}{argmax}q_{j} $ we know that $\mathbb{E}[\mathbb{E}[p_{i,n+1}|p_{i,n}]]\geq\mathbb{E}[p_{i,n}]$, meaning $\mathbb{E}[p_{i,n}]$ is an increasing sequence, and $\mathbb{E}[p_{i,0}] > 0 $, then it must be that $\underset{n\rightarrow\infty}{lim}\,\mathbb{E}[p_{i,n}]=1$. \\
Finally, since $\underset{j=1}{\overset{K}{\sum}}p_{j,\infty}=1$ and $p_{i,\infty}=1$ then:\\
$ \forall j\neq i:\, p_{j,\infty}=0$ \\
So we conclude that the probability to choose the arm with the highest expected reward converges to $1$, while the probability to choose any other arm converges to $0$ in this model. 

\paragraph{Combined Model} \mbox{ } \\
We first remind the reader the model:\\
\begin{gather*}
\mbox{For } i\in\{1,2,...,K\} \mbox{ and } 0 \leq \alpha \leq 1:\\ 
p_{i,n+1}=\begin{cases}
p_{i,n}+\alpha\cdot(1-p_{i,n}) & if\,\, j(n)=i\,\, and\,\, X_{i,n}=1\\
p_{i,n}-(1-\lambda)\cdot\alpha\cdot p_{i,n} & if\,\, j(n)=i\,\, and\,\, X_{i,n}=0\\
p_{i,n}-\alpha\cdot p_{i,n} & if\,\, j(n)\neq i\,\, and\,\, X_{j(n),n}=1\\
p_{i,n}+(1-\lambda)\cdot\alpha\cdot(1-p_{i,n}) & if\,\, j(n)\neq i\,\, and\,\, X_{j(n),n}=0 
\end{cases} 
\end{gather*}
We will compute the expected value of $p_{i,n+1}$ given $p_{i,n}$:
\begin{gather*}
\mathbb{E}[p_{i,n+1}|p_{i,n}] = p_{i,n} + \alpha \cdot \left( 1 - p_{i,n} \right)\cdot p_{i,n} \cdot q_i - \left( 1 - \lambda \right) \cdot \alpha \cdot p_{i,n} \cdot p_{i,n} \cdot (1 - q_i) \\
- \alpha \cdot p_{i,n} \cdot (1 - p_{i,n}) \cdot (1 - q_i) + \left( 1 - \lambda \right) \cdot \alpha \cdot \left( 1 - p_{i,n} \right) \cdot ( 1 - p_{i,n} ) \cdot q_i = \\
= \lambda \cdot\bigg(p_{i,n}+\alpha\cdot(1-p_{i,n})\cdot p_{i,n}\cdot q_{i}-\alpha\cdot p_{i,n}\cdot\underset{j\neq i}{\sum}p_{j,n}\cdot q_{j}\bigg) \\
+(1-\lambda)\cdot\bigg(p_{i,n}+\alpha\cdot(1-p_{i,n})\cdot p_{i,n}\cdot q_{i}-\alpha\cdot p_{i,n}\cdot\underset{j\neq i}{\sum}p_{j,n}\cdot q_{j} \\
-\alpha\cdot p_{i,n}\cdot p_{i,n}\cdot(1-q_{i})+\alpha\cdot(1-p_{i,n})\cdot\underset{j\neq i}{\sum}p_{j,n}\cdot(1-q_{j})\bigg) \\
= \lambda \cdot\bigg(p_{i,n}+\alpha\cdot(1-p_{i,n})\cdot p_{i,n}\cdot q_{i}-\alpha\cdot p_{i,n}\cdot\underset{j\neq i}{\sum}p_{j,n}\cdot q_{j}\bigg) \\
+(1-\lambda)\cdot\bigg(p_{i,n}+\alpha\cdot(1-p_{i,n})\cdot p_{i,n}\cdot q_{i}-\alpha\cdot p_{i,n}\cdot(1-p_{i,n})-\alpha\cdot p_{i,n}\cdot p_{i,n}\cdot(1-q_{i})+\alpha\cdot\underset{j\neq i}{\sum}p_{j,n}\cdot(1-q_{j})\bigg) \\
\end{gather*}
We now employ the fact that $K=2$ and get:
\begin{gather*}
\mathbb{E}[p_{i,n+1}|p_{i,n}] = \lambda \cdot\bigg(p_{i,n}+\alpha\cdot(1-p_{i,n})\cdot p_{i,n}\cdot q_{i}-\alpha\cdot p_{i,n}\cdot\underset{j\neq i}{\sum}p_{j,n}\cdot q_{j}\bigg) \\
+(1-\lambda)\cdot\bigg(p_{i,n}+\alpha\cdot(1-p_{i,n})\cdot p_{i,n}\cdot q_{i}-\alpha\cdot p_{i,n}\cdot(1-p_{i,n})-\alpha\cdot p_{i,n}\cdot p_{i,n}\cdot(1-q_{i})+\alpha\cdot(1-p_{i,n})\cdot q_{i}\bigg) \\
= \lambda \cdot\bigg(p_{i,n}+\alpha\cdot(1-p_{i,n})\cdot p_{i,n}\cdot q_{i}-\alpha\cdot p_{i,n}\cdot\underset{j\neq i}{\sum}p_{j,n}\cdot q_{j}\bigg) \\
+ (1-\lambda)\cdot\bigg(p_{i,n}+\alpha\cdot q_{i}\cdot(1-p_{i,n})-\alpha\cdot p_{i,n}\cdot(1-q_{i})\bigg)
\end{gather*}
So we got that the expected value of $p_{i,n+1}$ given $p_{i,n}$  is a linear interpolation between the expected value of it in the probability-matching model and the maximization model. \\
Thus we get:
\begin{gather*}
\mathbb{E}[p_{i,\infty}] = \lim_{n \rightarrow \infty}\mathbb{E}[p_{i,n+1}]=\lim_{n \rightarrow \infty}\mathbb{E}[\mathbb{E}[p_{i,n+1}|p_{i,n}]] \\
= \lim_{n \rightarrow \infty} \left( \lambda \cdot \mathbb{E}\left[ \bigg(p_{i,n}+\alpha\cdot(1-p_{i,n})\cdot p_{i,n}\cdot q_{i}-\alpha\cdot p_{i,n}\cdot\underset{j\neq i}{\sum}p_{j,n}\cdot q_{j}\bigg)\right] \right) \\
+ \lim_{n \rightarrow \infty} \left( (1-\lambda)\cdot \mathbb{E}\left[ \bigg(p_{i,n}+\alpha\cdot q_{i}\cdot(1-p_{i,n})-\alpha\cdot p_{i,n}\cdot(1-q_{i})\bigg) \right] \right)
\end{gather*}
So we have that the limit of this model is the interpolation of limits of the probability matching model and the maximization model.
Thus, if we denote by $p_{i,\infty}^1$ the convergent probability of the probability matching model, and similarly $p_{i,\infty}^2$ the convergent probability of the maximization model, then the convergence of this model will be:
\begin{gather*}
\mathbb{E}[p_{i,\infty}] =  \lambda \cdot p_{i,\infty}^2 + (1-\lambda) \cdot p_{i,\infty}^1
\end{gather*}

\subsection{Supplementary Figures}

\begin{figure}[H]
\centering
\includegraphics[scale=0.8]{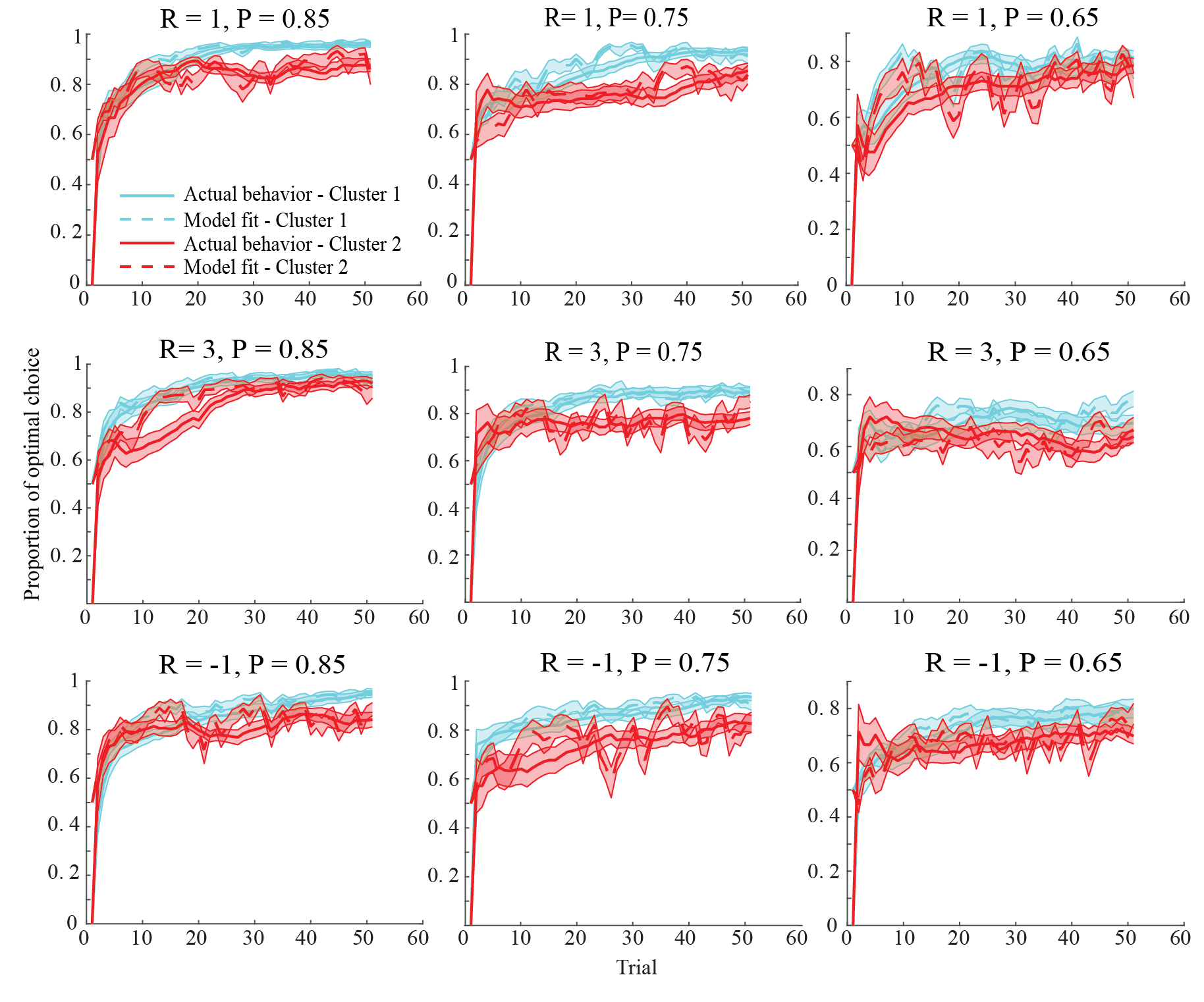}
\caption{Average learning performance of participants in the two clusters. The behavior of Cluster 1 is reminiscent of a maximization strategy, while Cluster 2 is more similar to a probability matching strategy.}
\label{suppFig:pm_supp4}
\end{figure}

\end{document}